\documentclass{article}

\PassOptionsToPackage{numbers, compress}{natbib}

\usepackage[preprint]{neurips_2024}

\usepackage[utf8]{inputenc} 
\usepackage[T1]{fontenc}    
\usepackage{hyperref}       
\usepackage{url}            
\usepackage{booktabs}       
\usepackage{amsfonts}       
\usepackage{nicefrac}       
\usepackage{xcolor}         
\usepackage{graphicx}

\usepackage{array}
\usepackage{multirow}

\title{From Human Memory to AI Memory: A Survey on Memory Mechanisms in the Era of LLMs}

\author{%
  Yaxiong Wu, Sheng Liang, Chen Zhang, Yichao Wang, Yongyue Zhang, \\ \textbf{Huifeng Guo, Ruiming Tang, Yong Liu} \\
  Huawei Noah’s Ark Lab \\
  \texttt{wu.yaxiong@huawei.com} \\
}

\begin{document}

\maketitle

\begin{abstract}
Memory is the process of encoding, storing, and retrieving information, allowing humans to retain experiences, knowledge, skills, and facts over time, and serving as the foundation for growth and effective interaction with the world. It plays a crucial role in shaping our identity, making decisions, learning from past experiences, building relationships, and adapting to changes. In the era of large language models (LLMs),  memory refers to the ability of an AI system to retain, recall, and use information from past interactions to improve future responses and interactions.  Although previous research and reviews have provided detailed descriptions of memory mechanisms, there is still a lack of a systematic review that summarizes and analyzes the relationship between the memory of LLM-driven AI systems and human memory, as well as how we can be inspired by human memory to construct more powerful memory systems. To achieve this, in this paper, we propose a comprehensive survey on the memory of LLM-driven AI systems. In particular, we first conduct a detailed analysis of the categories of human memory and relate them to the memory of AI systems. Second, we systematically organize existing memory-related work and propose a categorization method based on three dimensions (object, form, and time) and eight quadrants. Finally, we illustrate some open problems regarding the memory of current AI systems and outline possible future directions for memory in the era of large language models.

\end{abstract}

\section{Introduction}

Recently, large language models (LLMs) have become the core component of AI systems due to their powerful language understanding and generation capabilities, and are widely used in various applications such as intelligent customer service, automated writing, machine translation, information retrieval, and sentiment analysis~\cite{wang2024survey,li2024personal,zhao2023survey,chang2024survey}.
Unlike traditional AI systems, which rely on predefined rules and manually labeled features, LLM-driven AI systems offer greater flexibility, handling a diverse range of tasks with enhanced adaptability and contextual awareness.
Moreover, the introduction of memory enables LLMs to retain historical interactions with users and store contextual information, thereby providing more personalized, continuous, and context-aware responses in future interactions~\cite{li2024personal,chen2024all,zhang2025survey}.
AI systems powered by LLMs with memory capabilities will not only elevate the user experience but also support more complex and dynamic use cases, steering AI technology toward greater intelligence and human-centric design~\cite{zhang2024survey,jiang2024long}.

In neuroscience, human memory refers to the brain's ability to store, retain, and recall information~\cite{sherwood2004human,weng2023agent}. 
Human memory serves as the foundation for understanding the world, learning new knowledge, adapting to the environment, and making decisions, allowing us to preserve past experiences, skills, and knowledge, and helping us form our personal identity and behavior patterns~\cite{budson2023we}.
Human memory can be broadly classified into \textit{short-term memory} and \textit{long-term memory} based on the duration of new memory formation~\cite{baddeley2007working}. 
Short-term memory refers to the information we temporarily store and process, typically lasting from a few seconds to a few minutes, and includes sensory memory and working memory~\cite{budson2023we}. 
Long-term memory refers to the information we can store for extended periods, ranging from minutes to years, and includes declarative \textit{explicit memory} (such as episodic and semantic memory) and non-declarative \textit{implicit memory} (such as conditioned reflexes and procedural memory)~\cite{budson2023we}.
Human memory is a complex and dynamic process that relies on different memory systems to process information for various purposes, influencing how we understand and respond to the world. 
The different types of human memory and their working mechanisms can greatly inspire us to develop more scientific and reasonable memory-enhanced AI systems~\cite{gutierrez2024hipporag, applegarth2025exploring,gershman2025key, liu2025advanceschallengesfoundationagents}.

In the era of large language models (LLMs), the most typical memory-enhanced AI system is the LLM-powered autonomous agent system~\cite{weng2023agent}.
Large language model (LLM) powered agents are AI systems that can perform complex tasks using natural language, incorporating capabilities like planning, tool use, memory, and multi-step reasoning to enhance interactions and problem-solving~\cite{wang2024survey,li2024personal,weng2023agent}.
This memory-enhanced AI system is capable of autonomously decomposing complex tasks, remembering interaction history, and invoking and executing tools, thereby efficiently completing a series of intricate tasks.
In particular, memory, as a key component of the LLM-powered agent, can be defined as the process of acquiring, storing, retaining, and subsequently retrieving information~\cite{weng2023agent}. 
It enables the large language model to overcome the limitation of LLM's context window, allowing the agent to recall interaction history and make more accurate and intelligent decisions.
For instance, MemoryBank~\cite{zhong2024memorybank} proposed a long-term memory mechanism to allow LLMs for retrieving relevant memories, continuously evolving through continuous updates, and understanding and adapting to a user's personality by integrating information from previous interactions.
In addition, many commercial and open-source AI systems have also integrated memory systems to enhance the personalization capabilities of the system, such as OpenAI ChatGPT Memory~\cite{openai2024memory}, Apple Personal Context~\cite{apple2024personal}, mem0~\cite{mem0ai2024mem0}, MemoryScope~\cite{modelscope2024memoryscope}, etc.

Although previous studies and reviews have provided detailed explanations of memory mechanisms, most of the existing work focuses on analyzing and explaining memory from the temporal (time) dimension, specifically in terms of short-term and long-term memory~\cite{jiang2024long,zhang2024survey,zhong2024memorybank}.
We believe that categorizing memory solely based on the \textit{time} dimension is insufficient, as there are many other aspects (such as \textit{object} and \textit{form}) to memory classification in AI systems.
For example, from the object dimension, since AI systems often interact with humans, they need to perceive, store, recall, and use memories related to individual users, thus generating personal memories. Meanwhile, when AI systems perform complex tasks, they generate intermediate results (such as reasoning and planning processes, internet search results, etc.), which form system memory.
In addition, from the form dimension, since AI systems are powered by large language models (LLMs), they can store memories through the parametric memory encoded within the model parameters, as well as through non-parametric memory in the form of external memory documents that are stored and managed outside the model.
Therefore, insights that consider memory from the perspectives of object (personal and system), form (parametric and non-parametric), and time (short-term and long-term) are still lacking in the current era of large language models. 
There is still no comprehensive review that systematically analyzes the relationship between memory in LLM-driven AI systems and human memory, and how insights from human memory can be leveraged to build more efficient and powerful memory systems.

To fill this gap, this paper presents a comprehensive review of the memory mechanisms in LLM-driven AI systems.
First, we provide a detailed analysis of the categories of human memory and relate them to the memory systems in AI. 
In particular, we explore how human memory types — short-term memory (including sensory memory and working memory) and long-term memory (including explicit memory and implicit memory) — correspond to personal and system memory, parametric and non-parametric memory, and short-term and long-term memory in LLM-driven AI systems.
Next, we systematically organize the existing work related to memory and propose a classification method based on three dimensions (\textit{object}, \textit{form}, and \textit{time}) with eight quadrants.
In the object dimension, memory can be divided into personal memory and system memory; in the form dimension, it can be classified into parametric memory and non-parametric memory; in the time dimension, memory can be categorized into short-term memory and long-term memory.
Finally, based on the classification results from the three dimensions and eight quadrants mentioned above, we analyze some open issues in the memory of current AI systems and outline potential future directions for memory development in the era of large language models.

The main contributions of this paper are summarized as follows:
(1) We systematically and comprehensively define LLM-driven AI systems' memory and establish corresponding relationships with human memory.
(2) We propose a classification method for memory based on three dimensions (object, form, and time) and eight quadrants, which facilitates a more systematic exploration of memory in the era of large language models.
(3) From the perspective of enhancing personalized capabilities, we analyze and summarize research related to personal memory.
(4) From the perspective of AI system's ability to perform complex tasks, we analyze and summarize research related to system memory.
(5) We identify the existing issues and challenges in current memory research and point out potential future directions for development.

The remainder of the paper is organized as follows: 
In Section 2, we present a detailed description of human memory and AI systems' memory, comparing their differences and relationships, and introduce the classification method for memory based on three dimensions (object, form, and time) and eight quadrants.
In Section 3, we summarize research related to personal memory, aimed at enhancing the personalized response capabilities of AI systems.
In Section 4, we summarize research related to system memory, aimed at improving AI systems' ability to perform complex tasks.
In Section 5, we analyze some open issues related to memory and point out potential future directions for development.
Finally, in Section 6, we conclude the survey.

\section{Overview}

The human brain has evolved complex yet efficient memory mechanisms over a long period, enabling it to encode, store, and recall information effectively~\cite{sherwood2004human}. 
Accordingly, in the development of AI systems, we can draw insights from human memory to design effective \& efficient memory mechanisms or systems. 
In this section, we will first describe in detail the complex memory mechanisms and related memory systems of the human brain from the perspective of memory neuroscience. 
Then, we will discuss the memory mechanisms and types specific to LLM-driven AI systems. 
Finally, based on the memory features of LLM-driven AI systems, we will systematically review and categorize existing work from different dimensions.

\subsection{Human Memory}

Human memory typically relies on different memory systems to process information for various purposes, such as working memory for temporarily storing and processing information to support ongoing cognitive activities, and episodic memory for recording personal experiences and events for a long time~\cite{budson2023we}. 

\subsubsection{Short-Term and Long-Term Memory}
Based on the time range, human memory can be roughly divided into \textit{short-term memory} and \textit{long-term memory} according to the well-known Multi-Store Model (or Atkinson-Shiffrin Memory Model)~\cite{atkinson1968human}.

\paragraph{Short-Term Memory}
Short-term memory is a temporary storage system that holds small amounts of information for brief periods, typically ranging from seconds to minutes. 
It includes \textit{sensory memory}, which briefly captures raw sensory information from the environment (like sights or sounds), and \textit{working memory}, which actively processes and manipulates information to complete tasks such as problem-solving or learning. 
Together, these components allow humans to temporarily hold and work with information before either discarding it or transferring it to long-term memory.

\begin{itemize}
    \item \textbf{Sensory memory}: Sensory memory is the brief storage of sensory information we acquire from the external world, including iconic memory (visual), echoic memory (auditory), haptic memory (touch), and other sensory data. It typically lasts only a few milliseconds to a few seconds. Some sensory memories are transferred to working memory, while others are eventually stored in long-term memory (such as episodic memory).
    \item \textbf{Working memory}: Working memory is the system we use to temporarily store and process information. It not only helps us maintain current thoughts but also plays a role in decision-making and problem-solving. For example, when solving a math problem, it allows us to keep track of both the problem and the steps involved in finding the solution.
\end{itemize}

\paragraph{Long-Term Memory}
Long-term memory is a storage system that holds information for extended periods, ranging from minutes to a lifetime. 
It includes \textit{explicit memory}, which involves conscious recall of facts and events, and \textit{implicit memory}, which involves unconscious skills and habits, like riding a bike. 
These two types work together to help humans retain knowledge, experiences, and learned abilities over time.

\begin{itemize}
    \item \textbf{Explicit memory}: Explicit memory, also known as \textit{declarative memory}, refers to memories that we can easily verbalize or declare. It can be further divided into episodic memory and semantic memory. \textit{\textbf{Episodic memory}} refers to memories related to personal experiences and events, such as what you had for lunch. This type of memory is typically broken down into stages like encoding, storage, and retrieval. \textit{\textbf{Semantic memory}}, on the other hand, refers to memories related to facts and knowledge, such as knowing that the Earth is round or that the Earth orbits the Sun.
    \item \textbf{Implicit memory}: Implicit memory, also known as \textit{non-declarative memory}, refers to memories that are difficult to describe in words. It is associated with habits, skills, and procedures, and does not require conscious recall. \textit{\textbf{Procedural memory}} (or "muscle memory") is a typical form of implicit memory. It refers to memories gained through actions, such as riding a bicycle or playing the piano. The planning and coordination of movements are key components of procedural memory.
\end{itemize}

Multiple memory systems typically operate simultaneously, storing information in various ways across different brain regions. These memory systems are not completely independent; they interact with each other and, in many cases, depend on one another.
For example, when you hear a new song, the sensory memory in your ears and the brain regions responsible for processing sound will become active, storing the sound of the song for a few seconds. This sound is then transferred to your working memory system.
As you use your working memory and consciously think about the song, your episodic memory will automatically activate, recalling where you heard the song and what you were doing at the time.
As you hear the song in different places and at different times, a new semantic memory gradually forms, linking the melody of the song with its title. So, when you hear the song again, you'll remember the song's title, rather than a specific instance from your multiple listening experiences.
When you practice playing the song on the guitar, your procedural memory will remember the finger movements involved in playing the song.

\subsubsection{Memory Mechanisms}

Memory is the ability to encode, store and recall information. 
The three main processes involved in human memory are therefore \textit{encoding} (the process of acquiring and processing information into a form that can be stored), \textit{storage} (the retention of encoded information over time in short-term or long-term memory), and \textit{retrieval} (\textit{recall}, the process of accessing and bringing stored information back into conscious awareness when needed).

\begin{itemize}
    \item \textbf{Encoding} Memory encoding is the process of changing sensory information into a form that our brain can cope with and store effectively. In particular, there are different types of encoding in terms of how information is processed, such as \textit{visual encoding}, which involves processing information based on its visual features like color, shape, or texture; \textit{acoustic encoding}, which focuses on the auditory characteristics of information, such as pitch, tone, or rhythm; and \textit{semantic encoding}, which is based on the meaning of the information, making it easier to structure and remember. In addition, there are many approaches to make our brain better at encoding memory, such as \textit{mnemonics}, which involve using acronyms or peg-word systems to aid recall, \textit{chunking}, where information is broken down into smaller, meaningful units to enhance retention, \textit{imagination}, which strengthens encoding by linking images to words, and \textit{association}, where new information is connected to prior knowledge to improve understanding and long-term memory storage.
    \item \textbf{Storage} The storage of memory involves the coordinated activity of multiple brain regions, with key areas including: the \textit{prefrontal cortex}, which is associated with working memory and decision-making, helping us maintain and process information in the short term; the \textit{hippocampus}, which helps organize and consolidate information to form new explicit memories (such as episodic memory); the \textit{cerebral cortex}, which is involved in the storage and retrieval of semantic memory, allowing us to retain facts, concepts, and general knowledge over time; and the \textit{cerebellum}, which is primarily responsible for procedural memory formed through repetition.
    \item \textbf{Retrieval} Memory retrieval is the ability to access information and get it out of the memory storage. When we recall something, the brain reactivates neural pathways (also called synapses) linked to that memory. The prefrontal cortex helps in bringing memories back to awareness. Similarly, there are different types of memory retrieval, including \textit{recognition}, where we identify previously encountered information or stimuli, such as recognizing a familiar face or a fact we have learned before; \textit{recall}, which is the ability to retrieve information from memory without external cues, like remembering a phone number or address from memory; and \textit{relearning}, a process in which we reacquire previously learned but forgotten information, often at a faster pace than initial learning due to the residual memory traces that still exist.
\end{itemize}

In addition to the fundamental memory processing stages of encoding, storage, and retrieval, human memory also includes \textit{consolidation} (the process of stabilizing and strengthening memories to facilitate long-term storage), \textit{reconsolidation} (the modification or updating of previously stored memories when they are reactivated, allowing them to adapt to new information or contexts), \textit{reflection} (the active review and evaluation of one's memories to enhance self-awareness, improve learning strategies, and optimize decision-making), and \textit{forgetting} (the process by which information becomes inaccessible).

\begin{itemize}
    \item \textbf{Consolidation} Memory consolidation refers to the process of converting short-term memory into long-term memory, allowing information to be stably stored in the brain and reducing the likelihood of forgetting. It primarily involves the hippocampus and strengthens neural connections through \textit{synaptic plasticity} (strengthening of connections between neurons) and \textit{systems consolidation} (the gradual transfer and reorganization of memories from the hippocampus to the neocortex for long-term storage).
    \item \textbf{Reconsolidation} Memory reconsolidation refers to the process in which a previously stored memory is reactivated, entering an unstable state and requiring reconsolidation to maintain its storage. This process allows for the modification or updating of existing memories to adapt to new information or contexts, potentially leading to memory enhancement, weakening, or distortion. Once a memory is reactivated, it involves the hippocampus and amygdala and may be influenced by emotions, cognitive biases, or new information, resulting in memory adjustment or reshaping.
    \item \textbf{Reflection} Memory reflection refers to the process in which an individual actively reviews, evaluates, and examines their own memory content and processes to enhance self-awareness, adjust learning strategies, or optimize decision-making. It helps improve metacognitive ability, correct memory biases, facilitate deep learning, and regulate emotions. This process primarily relies on the brain's metacognitive ability (Metacognition) and involves the prefrontal cortex, which monitors and regulates memory functions.
    \item \textbf{Forgetting} Forgetting is a natural process that occurs when the brain fails to retrieve or retain information, which can result from \textit{encoding failure} (when information is not properly encoded due to lack of attention or meaningful connection), \textit{memory decay} (when memories fade over time without reinforcement as neural connections weaken), \textit{interference} (when similar or new memories compete with or overwrite existing ones), \textit{retrieval failure} (when information is inaccessible due to missing contextual cues despite being stored), or \textit{motivated forgetting} (when individuals consciously suppress or unconsciously repress traumatic or distressing memories). However, forgetting is a natural and necessary process that enables our brains to filter out irrelevant and outdated information, allowing us to prioritize what is most important for our current needs.
\end{itemize}

\subsection{Memory of LLM-driven AI Systems}

Similar to humans, LLM-driven AI systems also rely on memory systems to encode, store and recall information for future use. 
A typical example is the LLM-driven agent system, which leverages memory to enhance the agent system's abilities in reasoning, planning, personalization, and more~\cite{weng2023agent}.

\subsubsection{Fundamental Dimensions of AI Memory}

The memory of an LLM-driven AI system is closely related to the features of the LLM, that define how information is processed, stored, and retrieved based on its architecture and capabilities. 
We primarily categorize and organize memory based on three dimensions: \textit{object} (personal and system memory), \textit{form} (non-parametric and parametric memory), and \textit{time} (short-term and long-term memory).
These three dimensions comprehensively capture what type of information is retained (object), how information is stored (form), and how long it is preserved (time), aligning with both the functional structure of LLMs and practical requirements for efficient recall and adaptability.

\paragraph{Object Dimension}  The object dimension is closely tied to the interaction between LLM-driven AI systems and humans, as it defines how information is categorized based on its source and purpose. On one hand, the system receives human input and feedback (i.e., personal memory); on the other hand, it generates a series of intermediate output results during task execution (i.e., system memory). Personal memory helps the system improve its understanding of user behavior and enhances its personalization capabilities, while system memory can strengthen the system’s reasoning ability, such as in approaches like CoT (Chain-of-Thought)~\cite{wei2022chain} and ReAct~\cite{yao2022react}.

\paragraph{Form Dimension} The form dimension focuses on how memory is represented and stored in LLM-driven AI systems, shaping how information is encoded and retrieved. Some memory is embedded within the model’s parameters through training, forming parametric memory, while other memory exists externally in structured databases or retrieval mechanisms, constituting non-parametric memory. Non-parametric memory serves as a supplementary knowledge source that can be dynamically accessed by the large language model, enhancing its ability to retrieve relevant information in real-time, as seen in retrieval-augmented generation (RAG)~\cite{lewis2020retrieval}.

\paragraph{Time Dimension} The time dimension defines how long memory is retained and how it influences the LLM’s interactions over different timescales. Short-term memory refers to contextual information temporarily maintained within the current conversation, enabling coherence and continuity in multi-turn dialogues. In contrast, long-term memory consists of information from past interactions that is stored in an external database and retrieved when needed, allowing the model to retain user-specific knowledge and improve personalization over time. This distinction ensures that the system can balance real-time responsiveness with accumulated learning for enhanced adaptability.

In addition to the three primary dimensions discussed above, memory can also be classified based on other criteria, such as \textit{modality}, which distinguishes between unimodal memory (single data type) and multimodal memory (integrating multiple data types, such as text, images, and audio), or \textit{dynamics}, which differentiates between static memory (fixed and unchanging) and streaming memory (dynamically updated in real-time). However, these alternative classifications are not considered the primary criteria here, as our focus is on the core structural aspects that most directly influence memory organization and retrieval in LLM-driven AI systems.

\subsubsection{Parallels Between Human and AI Memory}

\begin{figure}[!htbp]
  \centering
  \includegraphics[width=0.98\columnwidth]{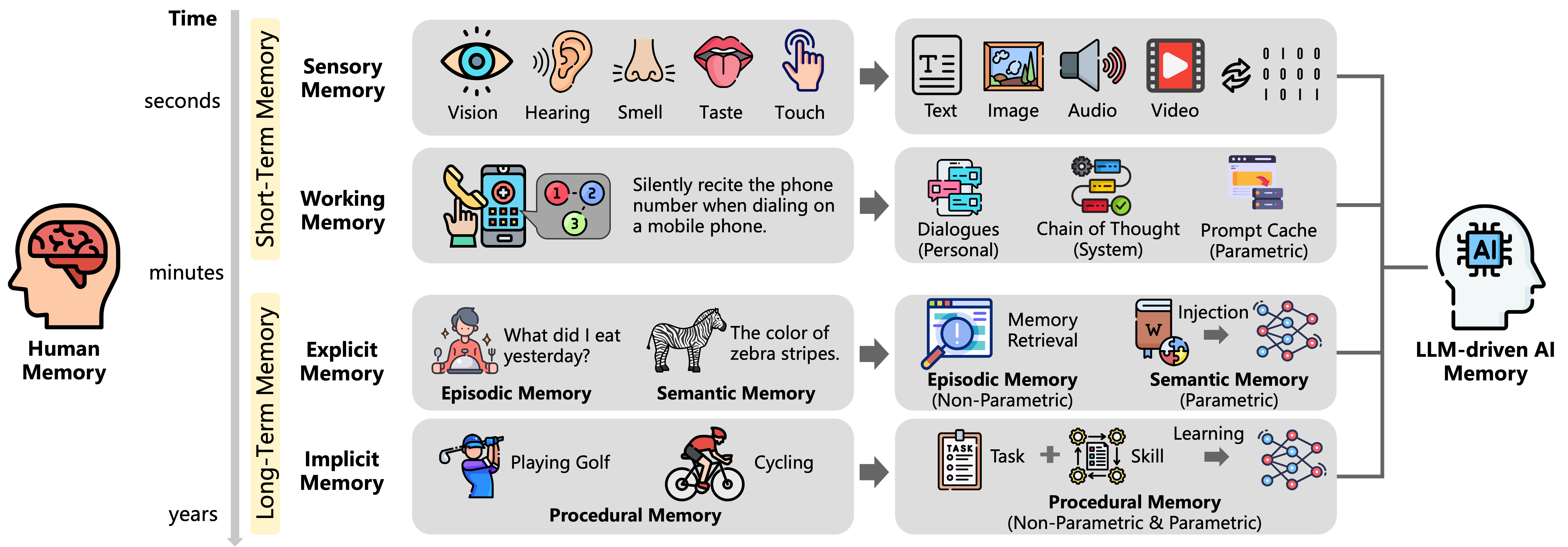}
  \caption{Illustrating the parallels between human and AI memory.}
  \label{fig:parallels}
\end{figure}

The memory of LLM-driven AI system exhibits similarities to human memory in terms of structure and function. Human memory is generally categorized into short-term memory and long-term memory, a distinction that also applies to AI memory systems. Below, we draw a direct comparison between these categories, mapping human cognitive memory processes to their counterparts in intelligent AI systems.
Figure~\ref{fig:parallels} illustrates the parallels between human and AI memory.

\begin{itemize}
    \item \textbf{Sensory Memory}: When an LLM-driven AI system perceives external information, it converts inputs such as text, images, speech, and video into machine-processable signals. This initial stage of information processing is analogous to human sensory memory, where raw data is briefly held before further cognitive processing. If these signals undergo additional processing, they transition into working memory, facilitating reasoning and decision-making. However, if no further processing or storage occurs, the information is quickly discarded, mirroring the transient nature of human sensory memory.
    \item \textbf{Working Memory}: The working memory of an AI system serves as a temporary storage and processing mechanism, enabling real-time reasoning and decision-making. It encompasses personal memory, such as contextual information retained during multi-turn dialogues, and system memory, including the chain of thoughts generated during task execution. As a form of short-term memory, working memory can undergo further processing and consolidation, eventually transitioning into long-term memory (e.g., episodic memory) that can be retrieved for future use. Additionally, during inference, large language models generate intermediate computational results, such as KV-Caches, which act as a form of parametric short-term memory that enhances efficiency by accelerating the inference process.
    \item \textbf{Explicit Memory}: The explicit memory of an AI system can be categorized into two distinct components. The first is non-parametric long-term memory, which involves the storage and retrieval of user-specific information, allowing the system to retain and utilize personalized data—analogous to episodic memory in humans. The second is parametric long-term memory, where factual knowledge and learned information are embedded within the model’s parameters, forming an internalized knowledge base—corresponding to semantic memory in human cognition. Together, these components enable the system to recall past interactions and apply acquired knowledge effectively.
    \item \textbf{Implicit Memory}: The implicit memory of an AI system encompasses the learned processes and patterns involved in task execution, enabling the development of specialized skills for specific tasks—analogous to procedural memory in humans. 
    This form of memory can parallel the human process of learning from both successes and failures in a non-parameterized manner, involving the reflection and refinement of accumulated traces, which allows the retention and replication of effective strategies from past experiences. 
    Additionally, it can be encoded within the model’s parameters, enabling the system to internalize task-related knowledge and perform operations efficiently without the need for explicit recall.
\end{itemize}

Beyond these parallels, insights from human memory can further guide the design of more effective and efficient AI memory systems, enhancing their ability to process, store, and retrieve information in a more structured and adaptive manner.

\subsubsection{3D-8Q Memory Taxonomy}

Building upon the three fundamental memory dimensions—object (personal \& system), form (non-parametric \& parametric), and time (short-term \& long-term)—as well as the established parallels between human and AI memory, we propose a \textit{three-dimensional, eight-quadrant (3D-8Q) memory taxonomy} for AI memory. 
This memory taxonomy systematically categorizes AI memory based on its function, storage mechanism, and retention duration, providing a structured approach to understanding and optimizing AI memory systems.
Table~\ref{tab:quadrants} presents the eight quadrants and their respective roles and functions.

\begin{table}[!htbp]
\centering
\newcolumntype{C}[1]{>{\centering\arraybackslash}m{#1}}
\label{tab:interaction}
\resizebox{0.98\textwidth}{!}{
\begin{tabular}{c|c|c|c|C{2.9cm}|>{\arraybackslash}p{6cm}}
\toprule
 \textbf{Object} & \textbf{Form} & \textbf{Time} & \textbf{Quadrant} & \textbf{Role} & \multicolumn{1}{c}{\textbf{Function}} \\ 
 \midrule
\multirow{13}{*}{Personal} & \multirow{7}{*}{Non-Parametric} & \multirow{3}{*}{Short-Term} & \multirow{3}{*}{I} & \multirow{3}{*}{Working Memory} & Supports real-time context supplementation, enhancing the AI's ability to maintain coherent interactions within a session. \\ \cline{3-6}
 & & \multirow{4}{*}{Long-Term} & \multirow{4}{*}{II} & \multirow{4}{*}{Episodic Memory} & Enables memory retention beyond session limits, allowing the system to recall and retrieve past user interactions for personalization. \\ 
 \cline{2-6}
 & \multirow{6}{*}{Parametric} & \multirow{3}{*}{Short-Term} & \multirow{3}{*}{III} & \multirow{3}{*}{Working Memory} & Temporarily enhances contextual understanding in ongoing interactions, improving response relevance and coherence. \\ \cline{3-6}
& & \multirow{3}{*}{Long-Term} & \multirow{3}{*}{IV} & \multirow{3}{*}{Semantic Memory} & Facilitates the continuous integration of newly acquired knowledge into the model, improving adaptability and personalization \\
\midrule
\multirow{15}{*}{System} & \multirow{7}{*}{Non-Parametric} & \multirow{3}{*}{Short-Term} & \multirow{3}{*}{V} & \multirow{3}{*}{Working Memory} & Assists in complex reasoning and decision-making by storing intermediate outputs such as chain-of-thought prompts. \\ \cline{3-6}
 & & \multirow{4}{*}{Long-Term} & \multirow{4}{*}{VI} & \multirow{4}{*}{Procedural Memory} & Captures historical experiences and self-reflection insights, enabling the AI to refine its reasoning and problem-solving skills over time. \\ 
 \cline{2-6}
 & \multirow{10}{*}{Parametric} & \multirow{5}{*}{Short-Term} & \multirow{5}{*}{VII} & \multirow{5}{*}{Working Memory} & Enhances computational efficiency through temporary parametric storage mechanisms such as KV-Caches, optimizing inference speed and reducing resource consumption. \\ \cline{3-6}
& & \multirow{5}{*}{Long-Term} & \multirow{5}{*}{VIII} &  \multirow{5}{=}{\centering{Semantic Memory Procedural Memory}} & Forms a foundational knowledge base encoded in the model’s parameters, serving as a long-term repository of factual \& conceptual knowledge and task-related knowledge. \\
\bottomrule
\end{tabular}}
\caption{Three-dimensional, eight-quadrant (3D-8Q) memory taxonomy for LLM-driven AI systems.}
\label{tab:quadrants}
\end{table}

Next, we will provide insights and descriptions of existing works from the perspectives of personal memory (in Section~\ref{sect:personal}) and system memory (in Section~\ref{sect:system}). In particualr, personal memory focuses more on the individual data perceived and observed by the model from the environment, while system memory emphasizes the system's internal or endogenous memory, such as the intermediate memory generated during task execution.

\section{Personal Memory}~\label{sect:personal}

Personal memory refers to the process of storing and utilizing human input and response data during interactions with an LLM-driven AI system. 
The development and application of personal memory play a crucial role in enhancing AI systems' personalization capabilities and improving user experience. 
In this section, we explore the concept of personal memory and relevant research, examining both non-parametric and parametric approaches to its construction and implementation.
Table~\ref{tab:personal} shows the categories, features, and related research work of personal memory.

\begin{table}[h!]
\centering
\newcolumntype{C}[1]{>{\centering\arraybackslash}m{#1}}
\resizebox{0.98\textwidth}{!}{
\begin{tabular}{C{1.5cm}|C{2.5cm}|C{2.5cm}|C{8.5cm}}
\toprule
 \textbf{Quadrant}& \textbf{Dimension} & \textbf{Feature} & \multicolumn{1}{c}{\textbf{Models}} \\ 
 \midrule
I & Personal Non-Parametric Short-Term & Multi-Turn Dialogue & ChatGPT~\cite{openai2022chatgpt}, DeepSeek-Chat~\cite{liu2024deepseek}, Claude~\cite{anthropic2023claude}, QWEN-CHAT~\cite{bai2023qwen}, Llama 2-Chat~\cite{touvron2023llama}, Gemini~\cite{team2023gemini}, PANGU-BOT~\cite{mi2022pangu}, ChatGLM~\cite{glm2024chatglm}, OpenAssistant~\cite{kopf2023openassistant}  \\
 \midrule
 \multirow{19}{*}{II} & \multirow{19}{=}{\centering{Personal Non-Parametric Long-Term}} & Personal Assistant & ChatGPT Memory~\cite{openai2024memory}, Apple Intelligence~\cite{apple2024personal}, Microsoft Recall~\cite{microsoft2024recall}, Me.bot~\cite{shang2024ai} \\ \cline{3-4}
 & & Open-Source Framework & MemoryScope~\cite{modelscope2024memoryscope}, mem0~\cite{mem0ai2024mem0}, Memary~\cite{memary2024memary}, LangGraph Memory~\cite{langchainai2024memory}, Charlie Mnemonic~\cite{goodai2024charlie}, Memobase~\cite{memodbio2025memobase}, Letta~\cite{letta2024letta}, Cognee~\cite{topoteretes2025cognee} \\ \cline{3-4}
 & & Construction & MPC~\cite{lee2023prompted}, RET-LLM~\cite{modarressi2023ret}, MemoryBank~\cite{zhong2024memorybank}, MemGPT~\cite{packer2023memgpt}, KGT~\cite{sun2024knowledge}, Evolving Conditional Memory~\cite{yuan2023personalized}, SECOM~\cite{pan2025memory}, Memory$^{3}$~\cite{yang2024memory3}, MemInsight~\cite{salama2025meminsight} \\ \cline{3-4}
 & & Management & MemoChat~\cite{lu2023memochat}, MemoryBank~\cite{zhong2024memorybank}, RMM~\cite{tan2025prospect}, LD-Agent~\cite{li2024hello}, A-MEM~\cite{xu2025mem}, Generative Agents~\cite{park2023generative}, EMG-RAG~\cite{wang2024crafting}, KGT~\cite{sun2024knowledge}, LLM-Rsum~\cite{wang2023recursively}, COMEDY~\cite{chen2024compress} \\ \cline{3-4}
 & & Retrieval & RET-LLM~\cite{modarressi2023ret}, ChatDB~\cite{hu2023chatdb}, Human-like Memory~\cite{hou2024my}, HippoRAG~\cite{gutierrez2024hipporag}, HippoRAG 2~\cite{gutierrez2025rag}, EgoRAG~\cite{yang2025egolife}, MemInsight~\cite{salama2025meminsight} \\ \cline{3-4}
 & & Usage & MemoCRS~\cite{xi2024memocrs}, RecMind~\cite{wang2023recmind}, RecAgent~\cite{wang2023recagent}, InteRecAgent~\cite{huang2023recommender}, SCM~\cite{wang2023enhancing}, ChatDev~\cite{qian2307chatdev}, MetaAgents~\cite{li2023metaagents}, S$^3$~\cite{gao2023s}, TradingGPT~\cite{li2023tradinggpt}, Memolet~\cite{yen2024memolet}, Synaptic Resonance~\cite{applegarth2025exploring}, MemReasoner~\cite{ko2024memreasoner} \\ \cline{3-4}
 & & Benchmark & MADial-Bench~\cite{he2024madial}, LOCOMO~\cite{maharana2024evaluating}, MemDaily~\cite{zhang2024memsim}, ChMapData~\cite{wu2025interpersonal}, MSC~\cite{xu-etal-2022-beyond}, MMRC~\cite{xue2025mmrc}, Ego4D~\cite{grauman2022ego4d}, EgoLife~\cite{yang2025egolife}, BABILong~\cite{kuratov2024babilong,kuratov2024search}  \\ 
 \midrule
  III & Personal Parametric Short-Term & Caching for Acceleration &  Prompt Cache~\cite{gim2024prompt}, Contextual Retrieval~\cite{anthropic2024contextual} \\ 
 \midrule
 IV & Personal Parametric Long-Term & Knowledge Editing & Character-LLM~\cite{shao2023character}, AI-Native Memory~\cite{shang2024ai}, MemoRAG~\cite{qian2024memorag}, Echo~\cite{liu2025echo} \\
\bottomrule
\end{tabular}}
\caption{Personal Memory}
\label{tab:personal}
\end{table}

\subsection{Contextual Personal Memory}

In personal memory, the non-parametric contextual memory that can be loaded is generally divided into two categories: the short-term memory of the current session's multi-turn dialogue and the long-term memory of historical dialogues across sessions.
The former can effectively supplement contextual information, while the latter can effectively fill in missing information and overcome the limitations of context length.

\subsubsection{Loading Multi-Turn Dialogue (Quadrant-I)}

In multi-turn dialogue scenarios, the conversation history of the current session can significantly enhance the LLM-driven AI system’s understanding of the user's real-time intent, leading to more relevant and contextually appropriate responses. 
Many modern dialogue systems are capable of handling multi-turn conversations and fully consider the current dialogue context in their responses. 
Notable examples include ChatGPT~\cite{openai2022chatgpt}, DeepSeek-Chat~\cite{liu2024deepseek}, and Claude~\cite{anthropic2023claude}, which excel at maintaining coherence and relevance over extended interactions.

For instance, ChatGPT~\cite{openai2022chatgpt} is a prime example of a multi-turn dialogue system where the conversation history of the current session serves as short-term memory, helping to supplement the contextual information of the dialogue. 
In ChatGPT, the dialogue memory is encoded in a role-content format, with distinct roles such as ``User'' and ``Assistant''.
This encoding allows the system to maintain clarity regarding the speaker and the flow of the conversation.

Through effective dialogue management at different levels, including ``Assistant'', ``Threads'', ``Messages'', and ``Runs'', the system can precisely track the state of each turn and each step of the conversation, ensuring continuity and consistency in interactions. 
Additionally, when the conversation length becomes too extensive, the dialogue system manages the conversation’s input by truncating the number of turns, thereby preventing the input from exceeding the model’s length limitations. 
This ensures that the system can continue processing the dialogue without losing track of essential context, maintaining the effectiveness of multi-turn interactions.

\subsubsection{Memory Retrieval-Augmented Generation (Quadrant-II)}

In cross-session dialogue scenarios, retrieving relevant user long-term memories from historical conversations can effectively supplement missing information in the current session, such as personal preferences and character relationships.
The advantage of memory retrieval-augmented generation is that large language models (LLMs) do not need to load all multi-session conversations. 
Given the limited length of LLMs' context windows—even when extended to millions of tokens—retrieving relevant information from historical sessions is also more efficient and cost-effective in terms of computation.
In addition to multi-session conversations, long-term personal memory also encompasses users’ behavioral history, preferences, and interaction records with AI agents over an extended period of time.

By leveraging retrieval-augmented generation from long-term memory, LLM-driven AI systems can better tailor their responses and behaviors, thereby improving user satisfaction and engagement.
For instance, a personal assistant that remembers a user's preferred news sources can prioritize those outlets in daily briefings, while a recommendation system that understands past viewing habits can suggest content more aligned with the user's tastes. 
Currently, many commercial and open-source platforms are striving to construct and utilize long-term memory for personalized AI systems—for example, ChatGPT Memory~\cite{openai2024memory} and Me.bot~\cite{shang2024ai} for personal assistants, and MemoryScope~\cite{modelscope2024memoryscope} and mem0~\cite{mem0ai2024mem0} as open-source frameworks. 
Long-term personal memory typically follows four core processing stages: \textit{construction}, \textit{management},\textit{ retrieval}, and \textit{usage}.
The second section of Table~\ref{tab:personal} (organized by rows) provides an overview of existing work on personal non-parametric long-term memory, classified based on their primary contributions.

\paragraph{Construction}
The construction of user memory requires extraction and refinement from raw memory data, such as multi-turn conversations. This process is analogous to human memory consolidation—the process of stabilizing and strengthening memories to facilitate their long-term storage.
Well-organized long-term memory enhances both the efficiency of storage and the effectiveness of retrieval in user memory. 
For example, MemoryBank~\cite{zhong2024memorybank} leverages a memory module to store conversation histories and summaries of key events, enabling the construction of a long-term user profile. 
Similarly, RET-LLM~\cite{modarressi2023ret} uses its memory module to retain essential factual knowledge about the external world, allowing the agent to monitor and update real-time environmental context relevant to the user.
In addition, to accommodate different types of memory, a variety of storage formats have been developed, including \textit{key-value}, \textit{graph}, and \textit{vector} representations. 
Specifically, \textit{key-value} formats~\cite{modarressi2023ret,salama2025meminsight,xi2024memocrs} enable efficient access to structured information such as user facts and preferences. 
\textit{Graph}-based formats~\cite{sun2024knowledge,gutierrez2024hipporag,gutierrez2025rag,mem0ai2024mem0} are designed to capture and represent relationships among entities, such as individuals and events. 
Meanwhile, \textit{vector} formats~\cite{zhong2024memorybank,pan2025memory,mem0ai2024mem0}, which are typically derived from textual, visual, or audio memory representations, are utilized to encode the semantic meaning and contextual information of conversations.
    
\paragraph{Management}
The management of user memory involves further processing and refinement of previously constructed memories, such as deduplication, merging, and conflict resolution. This process is analogous to human memory reconsolidation and reflection, where existing memories are reactivated, updated, and integrated to maintain coherence and relevance over time.
For instance, Reflective Memory Management (RMM)~\cite{tan2025prospect} is a user long-term memory management framework that combines Prospective Reflection for dynamic summarization with Retrospective Reflection for retrieval optimization via reinforcement learning. 
This dual-process approach addresses limitations such as rigid memory granularity and fixed retrieval mechanisms, enhancing the accuracy and flexibility of long-term memory management.
LD-Agent~\cite{li2024hello} enhances long-term dialogue personalization and consistency by constructing personalized persona information for both users and agents through a dynamic persona modeling module, while integrating retrieved memories to optimize response generation. 
A-MEM~\cite{xu2025mem} introduces a self-organizing memory system inspired by the Zettelkasten method~\cite{kadavy2021digital}, which constructs interconnected knowledge networks through dynamic indexing, linking, and memory evolution, enabling LLM agents to more flexibly organize, update, and retrieve long-term memories, thereby enhancing task adaptability and contextual awareness.
In addition, MemoryBank~\cite{zhong2024memorybank} incorporates a memory updating mechanism inspired by the Ebbinghaus Forgetting Curve~\cite{murre2015replication}, allowing the AI to forget or reinforce memories based on the time elapsed and their relative importance, thereby enabling a more human-like memory system and enhancing the user experience.

\paragraph{Retrieval}

Retrieving personal memory involves identifying memory entries relevant to the user's current request, and the retrieval method is closely tied to how the memory is stored. 
For key-value memory, ChatDB~\cite{hu2023chatdb} performs retrieval using SQL queries over structured databases. 
RET-LLM~\cite{modarressi2023ret}, on the other hand, employs a fuzzy search to retrieve triplet-structured memories, where information is stored as relationships between two entities connected by a predefined relation.
For graph-based memory, HippoRAG~\cite{gutierrez2024hipporag} constructs knowledge graphs over entities, phrases, and summarization to recall more relative and comprehensive memories, while HippoRAG 2~\cite{gutierrez2025rag} further combines original passages with phrase-based knowledge graphs to incorporate both conceptual and contextual information.
For vector memory, MemoryBank~\cite{zhong2024memorybank} adopts a dual-tower dense retrieval model, similar to Dense Passage Retrieval~\cite{karpukhin2020dense}, to accurately identify relevant memories. The resulting vector representations are then indexed using FAISS~\cite{johnson2019billion} for efficient similarity-based retrieval.
    
\paragraph{Usage}

The use of personal memory can effectively empower downstream applications with personalization, enhancing the user's individualized experience.
For instance, the recalled relevant memory is used as contextual information to enhance the personalized recommendation and response capability of the conversational recommender agents~\cite{xi2024memocrs,wang2023recmind,wang2023recagent,huang2023recommender}, improving the personalized user experience.
In addition to memory-augmented personalized dialogue and recommendation, personal memory can also be leveraged to enhance a wide range of applications, including software development~\cite{qian2307chatdev}, social-network simulation~\cite{li2023metaagents, gao2023s}, and financial trading~\cite{li2023tradinggpt}.

To facilitate in-depth research on personal memory, a variety of memory-related benchmarks have emerged in recent years, including long-term conversational memory (MADial-Bench~\cite{he2024madial}, LOCOMO~\cite{maharana2024evaluating}, MSC~\cite{xu-etal-2022-beyond}), everyday life memory (MemDaily~\cite{zhang2024memsim}), memory-aware proactive dialogue (ChMapData~\cite{wu2025interpersonal}), multimodal dialogue memory (MMRC~\cite{xue2025mmrc}), egocentric video understanding (Ego4D~\cite{grauman2022ego4d}, EgoLife~\cite{yang2025egolife}), and long-context reasoning-in-a-haystack (BABILong~\cite{kuratov2024babilong,kuratov2024search}).

\subsection{Parametric Personal Memory}

In addition to external non-parametric memory, a user's personal memory can also be stored parametrically. Specifically, personal data can be used to fine-tune an LLM, embedding the memory directly into its parameters (i.e., parametric long-term memory) to create a personalized LLM . Alternatively, historical dialogues can be cached as prompts during inference (i.e., parametric short-term memory), enabling quick reuse in future interactions.

\subsubsection{Memory Caching For Acceleration (Quadrant-III)}

Personal parametric short-term memory typically refers to intermediate attention states produced by the LLM when processing personal data, which is usually utilized as memory caches to accelerate inference.
Specifically, prompt caching~\cite{gim2024prompt} is usually used as an efficient data management technique that allows for the pre-storage of large amounts of personal data or information that may be frequently requested, such as a user's conversational history. 
For instance, during multi-turn dialogues, the dialogue system can quickly provide the personal context information directly from the parametric memory cache, avoiding the need to recalculate or retrieve it from the original data source, saving both time and resources.
Major platforms such as DeepSeek, Anthropic, OpenAI, and Google employ prompt caching to reduce API call costs and improve response speed in dialogue scenarios.
Moreover, personal parametric short-term memory can enhance the performance of retrieval-augmented generation (RAG) through Contextual Retrieval~\cite{anthropic2024contextual}, where prompt caching helps reduce the overhead of generating contextualized chunks.
At present, research specifically targeting caching techniques for personal memory data remains limited. Instead, most existing work considers caching as a fundamental capability of system memory, particularly in the context of key-value (KV) management and KV reuse. A more detailed discussion of these aspects is provided in Section~\ref{sect:system}.

\subsubsection{Personalized Knowledge Editing (Quadrant-IV)}

Personal parametric long-term memory utilizes personalized Knowledge Editing technology~\cite{wang2024knowledge}, such as Parameter-Efficient Fine-Tuning (PEFT)~\cite{han2024parameter}, to encode personal data into the LLM's parameters in a parametric manner, thereby facilitating the long-term, parameterized storage of memory. 
For instance, Character-LLM~\cite{shao2023character} enables the role-playing of specific characters, such as Beethoven, Queen Cleopatra, Julius Caesar, etc., by training large language models to remember the roles and experiences of these characters. 
AI-Native Memory~\cite{shang2024ai} proposes using deep neural network models, specifically large language models (LLMs), as Lifelong Personal Models (LPMs) to parameterize, compress, and continuously evolve personal memory through user interactions, enabling a more comprehensive understanding of the user.
MemoRAG~\cite{qian2024memorag} utilizes LLM parametric memory to store user conversation history and preferences, forming a personalized global memory that enhances personalization and enables tailored recommendations.
Echo~\cite{liu2025echo} is a large language model enhanced with temporal episodic memory, designed to improve performance in applications requiring multi-turn, complex memory-based dialogues.
The parameterization of personal long-term memory presents several challenges, notably the need to fine-tune models on individual user data, which demands substantial computational resources. This requirement significantly hinders the scalability and practical deployment of parametric approaches to long-term personal memory.

\subsection{Discussion}
In this section, we describe personal memory and related work from the perspectives of non-parametric and parametric approaches.
Specifically, personal non-parametric short-term memory necessitates efficient mechanisms for memory encoding and management. Existing literature predominantly emphasizes the design and implementation of systems that facilitate the construction, management, retrieval, and effective utilization of a user's personal non-parametric long-term memory. 
In contrast, personal parametric short-term memory can employ techniques such as prompt caching to reduce computational costs and enhance efficiency. 
Parametric long-term memory offers advantages in memory compression, thereby supporting a more comprehensive and global representation of the user’s accumulated experiences. 
Recent trends in the field indicate a growing interest in integrating both short-term and long-term memory paradigms, wherein parametric and non-parametric memory components complement and reinforce one another. 
The subsequent section will present a detailed discussion of system memory and its associated research developments.

\section{System Memory}~\label{sect:system}

System memory constitutes a critical component of LLM-driven AI systems. 
It encompasses a sequence of intermediate representations or results generated throughout the task execution process.
By leveraging system memory, LLM-driven AI systems can enhance their capabilities in reasoning, planning, and other higher-order cognitive functions. 
Moreover, the effective use of system memory contributes to the system’s capacity for self-evolution and continual improvement. 
In this section, we examine system memory and its associated research from both non-parametric and parametric perspectives.

\begin{table}[!htbp]
\centering
\newcolumntype{C}[1]{>{\centering\arraybackslash}m{#1}}
\resizebox{0.98\textwidth}{!}{
\begin{tabular}{C{1.5cm}|C{2.5cm}|C{2.5cm}|C{8.5cm}}
\toprule
 \textbf{Quadrant}& \textbf{Dimension} & \textbf{Feature} & \textbf{Models} \\ 
 \midrule
V & System Non-Parametric Short-Term & Reasoning \& Planning Enhancement & ReAct~\cite{yao2022react}, RAP~\cite{hao2023reasoning}, Reflexion~\cite{shinn2024reflexion}, Talker-Reasoner~\cite{christakopoulou2024agents}, TPTU~\cite{ruan2023tptu} \\
 \midrule
 VI & System Non-Parametric Long-Term & Reflection \& Refinement & Buffer of Thoughts~\cite{yang2024buffer}, AWM~\cite{wang2024agent}, Think-in-Memory~\cite{liu2023think}, GITM~\cite{zhu2023ghost}, Voyager~\cite{wang2023voyager}, Retroformer~\cite{yao2023retroformer}, Expel~\cite{zhao2024expel}, Synapse~\cite{zheng2023synapse}, MetaGPT~\cite{hong2023metagpt}, Learned Memory Bank~\cite{michelman2025enhancing}, M+~\cite{wang2025m+} \\
 \midrule
 \multirow{8}{*}{VII} & \multirow{8}{=}{\centering{System Parametric Short-Term}} & KV Management & LookupFFN~\cite{zeng2023lookupffn}, ChunkKV~\cite{liu2025chunkkv}, vLLM~\cite{kwon2023efficient}, FastServe~\cite{wu2023fast}, StreamingLLM~\cite{xiao2023efficient}, Orca~\cite{yu2022orca}, DistServe~\cite{zhong2024distserve}, LLM.int8()~\cite{dettmers2022gpt3}, FastGen~\cite{ge2023model}, Train Large, Then Compress~\cite{li2020train}, Scissorhands~\cite{liu2023scissorhands}, H$_2$O~\cite{zhang2023h2o}, Mooncake~\cite{qin2024mooncake}, MemServe~\cite{hu2024memserve}, SLM Serving~\cite{recasens2024towards}, IMPRESS~\cite{chen2025impress}, AdaServe~\cite{li2025adaserve}, MPIC~\cite{zhao2025mpic}, IntelLLM~\cite{li2024intelllm} \\ \cline{3-4}
 & & KV Reuse & KV Cache~\cite{pope2023efficiently}, Prompt Cache~\cite{gim2024prompt}, Contextual Retrieval~\cite{anthropic2024contextual}, CacheGen~\cite{liu2023cachegen}, ChunkAttention~\cite{ye2024chunkattention}, RAGCache~\cite{jin2024ragcache}, SGLang~\cite{zheng2024efficiently}, Ada-KV~\cite{feng2024ada}, HCache~\cite{gao2024fast}, Cake~\cite{jin2024compute}, EPIC~\cite{hu2024epic}, RelayAttention~\cite{zhu2024relayattention}, Marconi~\cite{pan2024marconi}, IKS~\cite{quinn2024accelerating}, FastCache~\cite{zhu2025fastcache}, Cache-Craft~\cite{agarwal2025cache}, KVLink~\cite{yang2025kvlink}, RAGServe~\cite{ray2024ragserve}, BumbleBee~\cite{kumaribumblebee} \\ 
 \midrule
 VIII & System Parametric Long-Term & Parametric Memory Structures & Memorizing Transformer~\cite{wu2022memorizing}, Focused Transformer~\cite{tworkowski2024focused}, MAC~\cite{tack2024online}, MemoryLLM~\cite{wang2024memoryllm}, WISE~\cite{wang2024wise}, LongMem~\cite{wang2023augmenting}, LM2~\cite{kanglm2}, Titans~\cite{behrouz2024titans} \\
\bottomrule
\end{tabular}}
\caption{System Memory}
\label{tab:system}
\end{table}

\subsection{Contextual System Memory}

From a temporal perspective, non-parametric short-term system memory refers to a series of reasoning and action results generated by large language models during task execution. 
This form of memory supports enhanced reasoning and planning within the context of the current task, thereby contributing to improved task accuracy, efficiency, and overall completion rates. 
In contrast, non-parametric long-term system memory represents a more abstracted and generalized form of short-term memory. 
It encompasses the consolidation of prior successful experiences and mechanisms of self-reflection based on historical interactions, which collectively facilitate the continual evolution and adaptive enhancement of LLM-driven AI systems.

\subsubsection{Reasoning \& Planning Enhancement (Quadrant-V)}

Analogous to human cognition, the reasoning and planning processes of large language models (LLMs) give rise to a sequence of short-term intermediate outputs. These outputs may reflect task-related attempts, which can be either successful or erroneous. Regardless of their correctness, such intermediate results serve as informative and constructive references that can guide subsequent task execution. This form of system non-parametric short-term memory plays a pivotal role in LLM-driven AI systems. Empirical evidence demonstrates that leveraging this memory structure significantly enhances the reasoning and planning capabilities of LLMs.
For instance, ReAct~\cite{yao2022react} integrates reasoning and action by generating intermediate reasoning steps alongside corresponding actions, enabling the model to alternate between thought and execution. This approach facilitates intelligent planning and adaptive decision-making in complex problem-solving scenarios. Similarly, Reflexion~\cite{shinn2024reflexion} introduces mechanisms for dynamic memory and self-reflection, allowing the LLM to self-evaluate and iteratively refine its behavior based on prior errors or limitations. This self-improvement loop promotes enhanced performance in future tasks, resembling a continuous learning and optimization process.

\subsubsection{Reflection \& Refinement (Quadrant-VI)}

The development of system non-parametric long-term memory parallels the human process of learning from both successes and failures. 
It involves the reflection upon and refinement of accumulated short-term memory traces. 
This memory mechanism enables the system not only to retain and replicate effective strategies from past experiences but also to extract valuable lessons from failures, thereby minimizing the likelihood of repeated errors. 
Through continuous updating and optimization, the system incrementally enhances its decision-making capabilities and improves its responsiveness to novel challenges. 
Moreover, the progressive accumulation of long-term memory empowers the system to address increasingly complex tasks with greater adaptability and resilience.
For instance, Buffer of Thoughts (BoT)~\cite{yang2024buffer} refines the chain of thoughts from historical tasks to form thought templates, which are then stored in a memory repository, guiding future reasoning and decision-making processes.
Agent Workflow Memory (AWM)~\cite{wang2024agent} introduces reusable paths, called workflows, and guides subsequent task generation by selecting different workflows.
Think-in-Memory (TiM)~\cite{liu2023think} continuously generates new thoughts based on conversation history, which is more conducive to reasoning and computation compared to raw observational data.
Ghost in the Minecraft (GITM)~\cite{zhu2023ghost} uses reference plans recorded in memory, allowing the agent planner to more efficiently handle encountered tasks, thereby improving task execution success rates.
Voyager~\cite{wang2023voyager} refines skills based on environmental feedback and stores acquired skills in memory, forming a skill library for future reuse in similar situations (e.g., fighting zombies vs. fighting spiders).
Retroformer~\cite{yao2023retroformer} leverages recent interaction trajectories as short-term memory and reflective feedback from past failures as long-term memory to guide decision-making and reasoning.
ExpeL~\cite{zhao2024expel} enhances task resolution by drawing on contextualized successful examples and abstracting insights from both successes and failures through comparative and pattern-based analysis of past experiences.

\subsection{Parametric System Memory}

The parametric system memory refers to the temporary storage of knowledge information in parametric forms, such as KV Cache~\cite{pope2023efficiently}, during the inference process (short-term memory), or the long-term editing and storage of knowledge information in the model parameters (long-term memory). 
The former, parametric short-term system memory, corresponds to human working memory, enabling cost reduction and efficiency improvement in large language model inference. 
The latter, parametric long-term system memory, corresponds to human semantic memory, facilitating the efficient integration of new knowledge.

\subsubsection{KV Management \& Reuse (Quadrant-VII)}

Parametric short-term system memory primarily focuses on the management and reuse of attention keys (Key) and values (Value) in LLMs, aiming to address issues such as high inference costs and latency during the reasoning process.
KV management optimizes memory efficiency and inference performance through techniques such as KV cache organization~\cite{kwon2023efficient}, compression~\cite{liu2025chunkkv}, and quantization~\cite{dettmers2022gpt3}.
In particular, vLLM~\cite{kwon2023efficient} is a high-efficiency LLM serving system built on PagedAttention, a virtual memory-inspired attention mechanism that enables near-zero KV cache waste and flexible sharing across requests, substantially improving batching efficiency and inference throughput.
ChunkKV~\cite{liu2025chunkkv} is a method for compressing the key-value cache in long-context inference with LLMs by grouping tokens into semantic chunks, retaining the most informative ones, and enabling layer-wise index reuse, thereby reducing memory and computational costs while outperforming existing approaches on several benchmarks.
LLM.int8()~\cite{dettmers2022gpt3} is a mixed-precision quantization method that combines vector-wise Int8 quantization with selective 16-bit handling of emergent outlier features, enabling memory-efficient inference of large language models (up to 175B parameters) without performance degradation.

Meanwhile, KV reuse focuses on reusing inference-related parameters through token-level KV Cache~\cite{pope2023efficiently} and sentence-level Prompt Cache~\cite{gim2024prompt}, which helps reduce computational costs and improve the efficiency of large language model (LLM) usage.
Specifically, KV Cache~\cite{pope2023efficiently} stores the attention keys (Key) and values (Value) generated by the neural network during sequence generation, allowing them to be reused in subsequent inference steps. This reuse accelerates attention computation in long-text generation and reduces redundant computation.
In contrast, Prompt Cache~\cite{gim2024prompt} operates at the sentence level by caching previous input prompts along with their corresponding output results. When similar prompts are encountered, the LLM can retrieve and return cached responses directly, saving computation and accelerating response generation.
By avoiding frequent recomputation of identical or similar contexts, KV reuse enables more efficient inference and significantly reduces computational overhead. 
Additionally, it enhances the flexibility and responsiveness of LLM-based systems in handling continuous or interactive tasks.
Building on these ideas, RAGCache~\cite{jin2024ragcache} introduces a multilevel dynamic caching system tailored for Retrieval-Augmented Generation (RAG), which caches intermediate knowledge states, optimizes memory replacement policies based on LLM inference and retrieval patterns, and overlaps retrieval with inference to significantly reduce latency and improve throughput.

Parametric short-term system memory overlaps somewhat with the previously mentioned parametric short-term personal memory in terms of technical approach. 
The difference lies in their focus: parametric short-term personal memory is more concerned with improving the processing of individual input data, while parametric short-term system memory focuses on optimizing the storage and reuse of system-level context during task execution. 
The former primarily addresses how to quickly process and adapt to an individual's input information, whereas the latter aims to reduce inference costs in multi-turn reasoning and enhance the consistency and efficiency of global tasks.

\subsubsection{Parametric Memory Structures (Quadrant-VIII)}

From the perspective of large language models (LLM) as long-term parametric memory, LLMs are not merely tools that provide immediate responses based on input and output; they can also store and integrate information over long time spans, forming an ever-evolving knowledge system.
LLMs based on the Transformer~\cite{vaswani2017attention} architecture are capable of memorizing knowledge information, primarily due to the self-attention mechanism in the Transformer-based model and the large-scale parameterized training approach.
By training on vast corpora, LLMs learn extensive world knowledge, language patterns, and solutions to various tasks. Additionally, LLMs can modify, update, or refine the internal knowledge through parameterized knowledge editing, allowing for more precise task handling or responses that better align with user needs.
MemoryLLM~\cite{wang2024memoryllm} has the ability to self-update and inject memory with new knowledge, effectively integrating new information and demonstrating excellent model editing performance and long-term information retention capabilities.
WISE~\cite{wang2024wise} is a lifelong editing framework for large language models that employs a dual-parametric memory design, with the main memory preserving pretrained knowledge and the side memory storing edited information.
It leverages a routing mechanism to dynamically access the appropriate memory during inference and uses knowledge sharding to distribute and integrate edits efficiently, ensuring reliability, generalization, and locality throughout continual updates.
The core function of parameterized knowledge editing~\cite{wang2024knowledge} is to enable large language models (LLMs) with dynamic and flexible knowledge updating capabilities, allowing them to respond to constantly changing task requirements, domain knowledge, and new information from the real world. 
This allows LLMs to remain efficient and accurate across various application scenarios and be customized and optimized according to user or environmental needs.

\subsection{Discussion}

In this section, we describe system memory and related work from the perspectives of non-parametric and parametric approaches.
Non-parametric short-term system memory can enhance the reasoning and planning abilities for current tasks, while non-parametric long-term system memory enables the reuse of successful experiences and the self-reflection based on historical experience, facilitating the evolution of LLM-driven AI system capabilities.
On the other hand, parametric short-term system memory can reduce costs and improve efficiency in large language model inference, and long-term parametric system memory can store and integrate information over long time spans, forming a continuously evolving knowledge system.
In the next section, we will summarize the issues and challenges in memory research in the era of large language models and point out potential future directions for development.

\section{Open Problems and Future Directions}

Although substantial progress has been made in current memory research across the three dimensions—object, form, and time—as well as within the eight corresponding quadrants, numerous open issues and challenges remain.
Building upon recent advancements and recognizing existing limitations, we outline the following promising directions for future research:

\paragraph{From Unimodal Memory to Multimodal Memory} 
In the era of large language models, LLM-driven AI systems are gradually expanding from being able to process only a single type of data (such as text) to handle multiple types of data simultaneously (such as text, images, audio, video, and even sensor data). 
This transition enhances perceptual capabilities and enables robust performance in complex real-world tasks.
For example, in the medical field, by combining text (medical records), images (medical imaging), and speech (doctor-patient conversations), AI systems can more accurately understand and diagnose medical conditions. 
Multimodal memory systems can integrate information from different sensory channels into a unified understanding, thereby approaching human cognitive processes more closely. 
Moreover, the expansion of multimodal memory also opens up possibilities for more personalized and interactive AI applications~\cite{zulfikar2024memoro}. 
For instance, personal AI assistants can not only communicate with users through text but also interpret users' emotions by recognizing facial expressions, voice intonations, or body language, thus providing more personalized and empathetic responses.

\paragraph{From Static Memory to Stream Memory}
Static memory can be viewed as a batch-processing approach to memory storage. It accumulates information or experiences in discrete batches, typically processing, storing, and retrieving them at specific intervals or predetermined points in time. As an offline memory model, static memory emphasizes the systematic organization and consolidation of large volumes of information, making it well-suited for long-term knowledge retention and structured learning.
In contrast, stream memory operates in a continuous, real-time manner. Analogous to data stream processing, it handles information as it arrives, prioritizing immediacy and adaptability. As an online or real-time memory model, stream memory focuses on the dynamic updating of information and rapid responsiveness to evolving contexts.
These two memory paradigms are not mutually exclusive and often function complementarily: while static memory supports the accumulation of stable, long-term knowledge, stream memory enables agile adaptation to ongoing tasks and real-time information demands.

\paragraph{From Specific Memory to Comprehensive Memory}
The human memory system comprises multiple interconnected subsystems—such as sensory memory, working memory, explicit memory, and implicit memory—each fulfilling distinct functions and contributing to the overall cognitive process.
In the context of large language models (LLMs), current memory architectures often concentrate on narrow or task-specific components, such as short-term memory for immediate inference or domain-specific knowledge storage.
While such targeted memory mechanisms can enhance performance in specific scenarios, their limited scope constrains the system’s overall flexibility, generalization, and adaptability.
Looking forward, the development of comprehensive and collaborative memory systems is essential. These systems should integrate diverse memory types and support efficient interaction, self-organization, and continual updating, enabling LLMs to manage increasingly complex and dynamic tasks.
By more closely emulating the multi-layered, multi-dimensional, and adaptive characteristics of human memory, such architectures have the potential to significantly advance the general intelligence and autonomy of LLM-based AI systems.

\paragraph{From Exclusive Memory to Shared Memory}
At present, the memory of each LLM-driven AI system operates independently, typically confined to a specific domain and tailored to processing isolated tasks or environments.
However, as AI technologies continue to evolve, memory systems are expected to become increasingly interconnected, transcending domain boundaries and enabling enhanced collaboration among models.
For instance, a large language model specialized in the medical domain could share its memory or knowledge base with another model focused on finance, facilitating cross-domain knowledge transfer and cooperative task solving.
Such a shared memory paradigm would not only improve the efficiency and adaptability of individual systems but also empower multiple LLMs to dynamically access and leverage one another’s domain-specific expertise.
This shift toward collaborative memory architectures could give rise to a more intelligent, resource-efficient network of AI systems capable of addressing complex, multi-domain challenges.
Ultimately, shared memory is poised to broaden the scope of AI applications and accelerate its integration into increasingly diverse and demanding real-world scenarios.

\paragraph{From Individual Privacy to Collective Privacy}
With the increasing prevalence of data sharing in the AI era, the focus of privacy protection is gradually shifting from the traditional notion of individual privacy to the broader and emerging concept of collective privacy.
Conventional privacy frameworks primarily aim to safeguard personal data, preventing unauthorized access, leakage, or misuse of individually identifiable information.
However, in the context of large language models, individual data is often aggregated into group-level datasets for large-scale analysis and prediction.
Collective privacy concerns the protection of the rights and interests of groups or communities whose data is used collectively, raising questions about how to prevent misuse, profiling, or excessive surveillance at the group level.
As memory systems in AI become more advanced and interconnected, ensuring collective privacy will emerge as a critical challenge.
Addressing this issue will require innovative techniques that can effectively balance the trade-off between data utility and privacy preservation~\cite{wang2025unveiling}.

\paragraph{From Rule-Based Evolution to Automated Evolution}
Traditional AI systems evolve by reflecting on past experiences—such as reusing successful strategies—based on accumulated knowledge and historical data.
However, this evolutionary process often depends on manually crafted rules and heuristic adjustments to enable such self-reflection.
While rule-based evolution can be effective, it inherently limits the system’s flexibility, scalability, and efficiency, with the quality and generalizability of the rules directly constraining the system’s adaptive capabilities.
Looking ahead, AI systems are expected to achieve automated evolution, dynamically adjusting and optimizing themselves by leveraging both personal and system-level memories in response to changing data and environmental contexts.
Such systems will be capable of autonomously identifying performance bottlenecks and initiating self-improvement without relying on explicit, human-defined rules.
This transition toward self-directed adaptation will significantly enhance system responsiveness, reduce the need for human intervention, and enable a more intelligent, dynamic, and continuously self-evolving paradigm.

\section{Conclusion}

Memory plays a pivotal role in the advancement of AI systems in the era of large language models (LLMs). It not only shapes the degree of personalization in AI behavior but also influences key capabilities such as adaptability, reasoning, planning, and self-evolution.
This article systematically examines the relationship between human memory and memory mechanisms in LLM-driven AI systems, exploring how principles of human cognition can inspire the design of more efficient and flexible memory architectures.
We begin by analyzing various categories of human memory—including perceptual memory, working memory, and long-term memory—and compare them with existing memory models in AI. Building upon this, we propose an eight-quadrant classification framework grounded in three dimensions: object, form, and time, offering a theoretical foundation for the construction of multi-level and comprehensive memory systems.
Furthermore, we review the current state of memory development in AI from both personal memory and system memory perspectives.
Finally, we identify key open challenges in contemporary AI memory design and outline promising directions for future research in the LLM era.
We believe that, with continued technological progress, AI systems will increasingly adopt more dynamic, adaptive, and intelligent memory architectures, thereby enabling more robust applications across complex, real-world tasks.

\bibliographystyle{unsrtnat}
\bibliography{main}

\begin{thebibliography}{100}

\bibitem{wang2024survey}
Lei Wang, Chen Ma, Xueyang Feng, Zeyu Zhang, Hao Yang, Jingsen Zhang, Zhiyuan Chen, Jiakai Tang, Xu~Chen, Yankai Lin, et~al.
\newblock A survey on large language model based autonomous agents.
\newblock {\em Frontiers of Computer Science}, 18(6):186345, 2024.

\bibitem{li2024personal}
Yuanchun Li, Hao Wen, Weijun Wang, Xiangyu Li, Yizhen Yuan, Guohong Liu, Jiacheng Liu, Wenxing Xu, Xiang Wang, Yi~Sun, et~al.
\newblock Personal llm agents: Insights and survey about the capability, efficiency and security.
\newblock {\em arXiv preprint arXiv:2401.05459}, 2024.

\bibitem{zhao2023survey}
Wayne~Xin Zhao, Kun Zhou, Junyi Li, Tianyi Tang, Xiaolei Wang, Yupeng Hou, Yingqian Min, Beichen Zhang, Junjie Zhang, Zican Dong, et~al.
\newblock A survey of large language models.
\newblock {\em arXiv preprint arXiv:2303.18223}, 1(2), 2023.

\bibitem{chang2024survey}
Yupeng Chang, Xu~Wang, Jindong Wang, Yuan Wu, Linyi Yang, Kaijie Zhu, Hao Chen, Xiaoyuan Yi, Cunxiang Wang, Yidong Wang, et~al.
\newblock A survey on evaluation of large language models.
\newblock {\em ACM transactions on intelligent systems and technology}, 15(3):1--45, 2024.

\bibitem{chen2024all}
Bo~Chen, Xinyi Dai, Huifeng Guo, Wei Guo, Weiwen Liu, Yong Liu, Jiarui Qin, Ruiming Tang, Yichao Wang, Chuhan Wu, et~al.
\newblock All roads lead to rome: Unveiling the trajectory of recommender systems across the llm era.
\newblock {\em arXiv preprint arXiv:2407.10081}, 2024.

\bibitem{zhang2025survey}
Chen Zhang, Xinyi Dai, Yaxiong Wu, Qu~Yang, Yasheng Wang, Ruiming Tang, and Yong Liu.
\newblock A survey on multi-turn interaction capabilities of large language models.
\newblock {\em arXiv preprint arXiv:2501.09959}, 2025.

\bibitem{zhang2024survey}
Zeyu Zhang, Xiaohe Bo, Chen Ma, Rui Li, Xu~Chen, Quanyu Dai, Jieming Zhu, Zhenhua Dong, and Ji-Rong Wen.
\newblock A survey on the memory mechanism of large language model based agents.
\newblock {\em arXiv preprint arXiv:2404.13501}, 2024.

\bibitem{jiang2024long}
Xun Jiang, Feng Li, Han Zhao, Jiaying Wang, Jun Shao, Shihao Xu, Shu Zhang, Weiling Chen, Xavier Tang, Yize Chen, et~al.
\newblock Long term memory: The foundation of ai self-evolution.
\newblock {\em arXiv preprint arXiv:2410.15665}, 2024.

\bibitem{sherwood2004human}
Lauralee Sherwood, Robert~Thomas Kell, and Christopher Ward.
\newblock {\em Human physiology: from cells to systems}.
\newblock Thomson/Brooks/Cole, 2004.

\bibitem{weng2023agent}
Lilian Weng.
\newblock Llm-powered autonomous agents.
\newblock {\em lilianweng.github.io}, Jun 2023.

\bibitem{budson2023we}
Andrew~E Budson and Elizabeth~A Kensinger.
\newblock {\em Why we forget and how to remember better: the science behind memory}.
\newblock Oxford University Press, 2023.

\bibitem{baddeley2007working}
Alan Baddeley.
\newblock {\em Working memory, thought, and action}, volume~45.
\newblock OuP Oxford, 2007.

\bibitem{gutierrez2024hipporag}
Bernal~Jim{\'e}nez Guti{\'e}rrez, Yiheng Shu, Yu~Gu, Michihiro Yasunaga, and Yu~Su.
\newblock Hipporag: Neurobiologically inspired long-term memory for large language models.
\newblock {\em arXiv preprint arXiv:2405.14831}, 2024.

\bibitem{applegarth2025exploring}
George Applegarth, Christian Weatherstone, Maximilian Hollingsworth, Henry Middlebrook, and Marcus Irvin.
\newblock Exploring synaptic resonance in large language models: A novel approach to contextual memory integration.
\newblock {\em arXiv preprint arXiv:2502.10699}, 2025.

\bibitem{gershman2025key}
Samuel~J Gershman, Ila Fiete, and Kazuki Irie.
\newblock Key-value memory in the brain.
\newblock {\em arXiv preprint arXiv:2501.02950}, 2025.

\bibitem{liu2025advanceschallengesfoundationagents}
Bang Liu, Xinfeng Li, Jiayi Zhang, Jinlin Wang, Tanjin He, Sirui Hong, Hongzhang Liu, Shaokun Zhang, Kaitao Song, Kunlun Zhu, Yuheng Cheng, Suyuchen Wang, Xiaoqiang Wang, Yuyu Luo, Haibo Jin, Peiyan Zhang, Ollie Liu, Jiaqi Chen, Huan Zhang, Zhaoyang Yu, Haochen Shi, Boyan Li, Dekun Wu, Fengwei Teng, Xiaojun Jia, Jiawei Xu, Jinyu Xiang, Yizhang Lin, Tianming Liu, Tongliang Liu, Yu~Su, Huan Sun, Glen Berseth, Jianyun Nie, Ian Foster, Logan Ward, Qingyun Wu, Yu~Gu, Mingchen Zhuge, Xiangru Tang, Haohan Wang, Jiaxuan You, Chi Wang, Jian Pei, Qiang Yang, Xiaoliang Qi, and Chenglin Wu.
\newblock Advances and challenges in foundation agents: From brain-inspired intelligence to evolutionary, collaborative, and safe systems, 2025.

\bibitem{zhong2024memorybank}
Wanjun Zhong, Lianghong Guo, Qiqi Gao, He~Ye, and Yanlin Wang.
\newblock Memorybank: Enhancing large language models with long-term memory.
\newblock In {\em Proceedings of the AAAI Conference on Artificial Intelligence}, volume~38, pages 19724--19731, 2024.

\bibitem{openai2024memory}
OpenAI.
\newblock Memory and new controls for chatgpt.
\newblock {\em openai.com}, February 2024.

\bibitem{apple2024personal}
Apple.
\newblock Introducing apple intelligence, the personal intelligence system that puts powerful generative models at the core of iphone, ipad, and mac.
\newblock {\em apple.com}, June 2024.

\bibitem{mem0ai2024mem0}
mem0ai.
\newblock mem0: The memory layer for personalized ai.
\newblock {\em mem0.ai}, July 2024.

\bibitem{modelscope2024memoryscope}
ModelScope.
\newblock Memoryscope: Equip your llm chatbot with a powerful and flexible long term memory system.
\newblock {\em github.com}, September 2024.

\bibitem{atkinson1968human}
Richard~C Atkinson and Richard~M Shiffrin.
\newblock Human memory: A proposed system and its control processes.
\newblock In {\em Psychology of learning and motivation}, volume~2, pages 89--195. Elsevier, 1968.

\bibitem{wei2022chain}
Jason Wei, Xuezhi Wang, Dale Schuurmans, Maarten Bosma, Fei Xia, Ed~Chi, Quoc~V Le, Denny Zhou, et~al.
\newblock Chain-of-thought prompting elicits reasoning in large language models.
\newblock {\em Advances in neural information processing systems}, 35:24824--24837, 2022.

\bibitem{yao2022react}
Shunyu Yao, Jeffrey Zhao, Dian Yu, Nan Du, Izhak Shafran, Karthik Narasimhan, and Yuan Cao.
\newblock React: Synergizing reasoning and acting in language models.
\newblock {\em arXiv preprint arXiv:2210.03629}, 2022.

\bibitem{lewis2020retrieval}
Patrick Lewis, Ethan Perez, Aleksandra Piktus, Fabio Petroni, Vladimir Karpukhin, Naman Goyal, Heinrich K{\"u}ttler, Mike Lewis, Wen-tau Yih, Tim Rockt{\"a}schel, et~al.
\newblock Retrieval-augmented generation for knowledge-intensive nlp tasks.
\newblock {\em Advances in neural information processing systems}, 33:9459--9474, 2020.

\bibitem{openai2022chatgpt}
OpenAI.
\newblock Introducing chatgpt.
\newblock {\em openai.com}, November 2022.

\bibitem{liu2024deepseek}
Aixin Liu, Bei Feng, Bing Xue, Bingxuan Wang, Bochao Wu, Chengda Lu, Chenggang Zhao, Chengqi Deng, Chenyu Zhang, Chong Ruan, et~al.
\newblock Deepseek-v3 technical report.
\newblock {\em arXiv preprint arXiv:2412.19437}, 2024.

\bibitem{anthropic2023claude}
Anthropic.
\newblock Introducing claude.
\newblock {\em anthropic.com}, March 2023.

\bibitem{bai2023qwen}
Jinze Bai, Shuai Bai, Yunfei Chu, Zeyu Cui, Kai Dang, Xiaodong Deng, Yang Fan, Wenbin Ge, Yu~Han, Fei Huang, et~al.
\newblock Qwen technical report.
\newblock {\em arXiv preprint arXiv:2309.16609}, 2023.

\bibitem{touvron2023llama}
Hugo Touvron, Louis Martin, Kevin Stone, Peter Albert, Amjad Almahairi, Yasmine Babaei, Nikolay Bashlykov, Soumya Batra, Prajjwal Bhargava, Shruti Bhosale, et~al.
\newblock Llama 2: Open foundation and fine-tuned chat models.
\newblock {\em arXiv preprint arXiv:2307.09288}, 2023.

\bibitem{team2023gemini}
Gemini Team, Rohan Anil, Sebastian Borgeaud, Jean-Baptiste Alayrac, Jiahui Yu, Radu Soricut, Johan Schalkwyk, Andrew~M Dai, Anja Hauth, Katie Millican, et~al.
\newblock Gemini: a family of highly capable multimodal models.
\newblock {\em arXiv preprint arXiv:2312.11805}, 2023.

\bibitem{mi2022pangu}
Fei Mi, Yitong Li, Yulong Zeng, Jingyan Zhou, Yasheng Wang, Chuanfei Xu, Lifeng Shang, Xin Jiang, Shiqi Zhao, and Qun Liu.
\newblock Pangu-bot: Efficient generative dialogue pre-training from pre-trained language model.
\newblock {\em arXiv preprint arXiv:2203.17090}, 2022.

\bibitem{glm2024chatglm}
Team GLM, Aohan Zeng, Bin Xu, Bowen Wang, Chenhui Zhang, Da~Yin, Dan Zhang, Diego Rojas, Guanyu Feng, Hanlin Zhao, et~al.
\newblock Chatglm: A family of large language models from glm-130b to glm-4 all tools.
\newblock {\em arXiv preprint arXiv:2406.12793}, 2024.

\bibitem{kopf2023openassistant}
Andreas K{\"o}pf, Yannic Kilcher, Dimitri Von~R{\"u}tte, Sotiris Anagnostidis, Zhi~Rui Tam, Keith Stevens, Abdullah Barhoum, Duc Nguyen, Oliver Stanley, Rich{\'a}rd Nagyfi, et~al.
\newblock Openassistant conversations-democratizing large language model alignment.
\newblock {\em Advances in Neural Information Processing Systems}, 36:47669--47681, 2023.

\bibitem{microsoft2024recall}
Microsoft.
\newblock Recall overview.
\newblock {\em microsoft.com}, February 2025.

\bibitem{shang2024ai}
Jingbo Shang, Zai Zheng, Jiale Wei, Xiang Ying, Felix Tao, and Mindverse Team.
\newblock Ai-native memory: A pathway from llms towards agi.
\newblock {\em arXiv preprint arXiv:2406.18312}, 2024.

\bibitem{memary2024memary}
Memary.
\newblock Beyond short-term memory: How memary makes chatbots remember.
\newblock {\em github.com}, April 2024.

\bibitem{langchainai2024memory}
langchain ai.
\newblock Langgraph memory service.
\newblock {\em github.com}, October 2024.

\bibitem{goodai2024charlie}
GoodAI.
\newblock Charlie mnemonic.
\newblock {\em github.com}, March 2024.

\bibitem{memodbio2025memobase}
memodb io.
\newblock Memobase: User profile-based memory for genai apps.
\newblock {\em memobase.io}, January 2025.

\bibitem{letta2024letta}
Letta-AI.
\newblock Letta.
\newblock {\em github.com}, September 2024.

\bibitem{topoteretes2025cognee}
Cognee.ai.
\newblock Cognee.
\newblock {\em github.com}, October 2024.

\bibitem{lee2023prompted}
Gibbeum Lee, Volker Hartmann, Jongho Park, Dimitris Papailiopoulos, and Kangwook Lee.
\newblock Prompted llms as chatbot modules for long open-domain conversation.
\newblock {\em arXiv preprint arXiv:2305.04533}, 2023.

\bibitem{modarressi2023ret}
Ali Modarressi, Ayyoob Imani, Mohsen Fayyaz, and Hinrich Sch{\"u}tze.
\newblock Ret-llm: Towards a general read-write memory for large language models.
\newblock {\em arXiv preprint arXiv:2305.14322}, 2023.

\bibitem{packer2023memgpt}
Charles Packer, Sarah Wooders, Kevin Lin, Vivian Fang, Shishir~G Patil, Ion Stoica, and Joseph~E Gonzalez.
\newblock Memgpt: Towards llms as operating systems.
\newblock {\em arXiv preprint arXiv:2310.08560}, 2023.

\bibitem{sun2024knowledge}
Jingwei Sun, Zhixu Du, and Yiran Chen.
\newblock Knowledge graph tuning: Real-time large language model personalization based on human feedback.
\newblock {\em arXiv preprint arXiv:2405.19686}, 2024.

\bibitem{yuan2023personalized}
Ruifeng Yuan, Shichao Sun, Yongqi Li, Zili Wang, Ziqiang Cao, and Wenjie Li.
\newblock Personalized large language model assistant with evolving conditional memory.
\newblock {\em arXiv preprint arXiv:2312.17257}, 2023.

\bibitem{pan2025memory}
Zhuoshi Pan, Qianhui Wu, Huiqiang Jiang, Xufang Luo, Hao Cheng, Dongsheng Li, Yuqing Yang, Chin-Yew Lin, H~Vicky Zhao, Lili Qiu, et~al.
\newblock On memory construction and retrieval for personalized conversational agents.
\newblock {\em arXiv preprint arXiv:2502.05589}, 2025.

\bibitem{yang2024memory3}
Hongkang Yang, Zehao Lin, Wenjin Wang, Hao Wu, Zhiyu Li, Bo~Tang, Wenqiang Wei, Jinbo Wang, Zeyun Tang, Shichao Song, et~al.
\newblock Memory3: Language modeling with explicit memory.
\newblock {\em arXiv preprint arXiv:2407.01178}, 2024.

\bibitem{salama2025meminsight}
Rana Salama, Jason Cai, Michelle Yuan, Anna Currey, Monica Sunkara, Yi~Zhang, and Yassine Benajiba.
\newblock Meminsight: Autonomous memory augmentation for llm agents.
\newblock {\em arXiv preprint arXiv:2503.21760}, 2025.

\bibitem{lu2023memochat}
Junru Lu, Siyu An, Mingbao Lin, Gabriele Pergola, Yulan He, Di~Yin, Xing Sun, and Yunsheng Wu.
\newblock Memochat: Tuning llms to use memos for consistent long-range open-domain conversation.
\newblock {\em arXiv preprint arXiv:2308.08239}, 2023.

\bibitem{tan2025prospect}
Zhen Tan, Jun Yan, I~Hsu, Rujun Han, Zifeng Wang, Long~T Le, Yiwen Song, Yanfei Chen, Hamid Palangi, George Lee, et~al.
\newblock In prospect and retrospect: Reflective memory management for long-term personalized dialogue agents.
\newblock {\em arXiv preprint arXiv:2503.08026}, 2025.

\bibitem{li2024hello}
Hao Li, Chenghao Yang, An~Zhang, Yang Deng, Xiang Wang, and Tat-Seng Chua.
\newblock Hello again! llm-powered personalized agent for long-term dialogue.
\newblock {\em arXiv preprint arXiv:2406.05925}, 2024.

\bibitem{xu2025mem}
Wujiang Xu, Zujie Liang, Kai Mei, Hang Gao, Juntao Tan, and Yongfeng Zhang.
\newblock A-mem: Agentic memory for llm agents.
\newblock {\em arXiv preprint arXiv:2502.12110}, 2025.

\bibitem{park2023generative}
Joon~Sung Park, Joseph O'Brien, Carrie~Jun Cai, Meredith~Ringel Morris, Percy Liang, and Michael~S Bernstein.
\newblock Generative agents: Interactive simulacra of human behavior.
\newblock In {\em Proceedings of the 36th annual acm symposium on user interface software and technology}, pages 1--22, 2023.

\bibitem{wang2024crafting}
Zheng Wang, Zhongyang Li, Zeren Jiang, Dandan Tu, and Wei Shi.
\newblock Crafting personalized agents through retrieval-augmented generation on editable memory graphs.
\newblock {\em arXiv preprint arXiv:2409.19401}, 2024.

\bibitem{wang2023recursively}
Qingyue Wang, Liang Ding, Yanan Cao, Zhiliang Tian, Shi Wang, Dacheng Tao, and Li~Guo.
\newblock Recursively summarizing enables long-term dialogue memory in large language models.
\newblock {\em arXiv preprint arXiv:2308.15022}, 2023.

\bibitem{chen2024compress}
Nuo Chen, Hongguang Li, Juhua Huang, Baoyuan Wang, and Jia Li.
\newblock Compress to impress: Unleashing the potential of compressive memory in real-world long-term conversations.
\newblock {\em arXiv preprint arXiv:2402.11975}, 2024.

\bibitem{hu2023chatdb}
Chenxu Hu, Jie Fu, Chenzhuang Du, Simian Luo, Junbo Zhao, and Hang Zhao.
\newblock Chatdb: Augmenting llms with databases as their symbolic memory.
\newblock {\em arXiv preprint arXiv:2306.03901}, 2023.

\bibitem{hou2024my}
Yuki Hou, Haruki Tamoto, and Homei Miyashita.
\newblock " my agent understands me better": Integrating dynamic human-like memory recall and consolidation in llm-based agents.
\newblock In {\em Extended Abstracts of the CHI Conference on Human Factors in Computing Systems}, pages 1--7, 2024.

\bibitem{gutierrez2025rag}
Bernal~Jim{\'e}nez Guti{\'e}rrez, Yiheng Shu, Weijian Qi, Sizhe Zhou, and Yu~Su.
\newblock From rag to memory: Non-parametric continual learning for large language models.
\newblock {\em arXiv preprint arXiv:2502.14802}, 2025.

\bibitem{yang2025egolife}
Jingkang Yang, Shuai Liu, Hongming Guo, Yuhao Dong, Xiamengwei Zhang, Sicheng Zhang, Pengyun Wang, Zitang Zhou, Binzhu Xie, Ziyue Wang, et~al.
\newblock Egolife: Towards egocentric life assistant.
\newblock {\em arXiv preprint arXiv:2503.03803}, 2025.

\bibitem{xi2024memocrs}
Yunjia Xi, Weiwen Liu, Jianghao Lin, Bo~Chen, Ruiming Tang, Weinan Zhang, and Yong Yu.
\newblock Memocrs: Memory-enhanced sequential conversational recommender systems with large language models.
\newblock In {\em Proceedings of the 33rd ACM International Conference on Information and Knowledge Management}, pages 2585--2595, 2024.

\bibitem{wang2023recmind}
Yancheng Wang, Ziyan Jiang, Zheng Chen, Fan Yang, Yingxue Zhou, Eunah Cho, Xing Fan, Xiaojiang Huang, Yanbin Lu, and Yingzhen Yang.
\newblock Recmind: Large language model powered agent for recommendation.
\newblock {\em arXiv preprint arXiv:2308.14296}, 2023.

\bibitem{wang2023recagent}
Lei Wang, Jingsen Zhang, Xu~Chen, Yankai Lin, Ruihua Song, Wayne~Xin Zhao, and Ji-Rong Wen.
\newblock Recagent: A novel simulation paradigm for recommender systems.
\newblock {\em arXiv preprint arXiv:2306.02552}, 2023.

\bibitem{huang2023recommender}
Xu~Huang, Jianxun Lian, Yuxuan Lei, Jing Yao, Defu Lian, and Xing Xie.
\newblock Recommender ai agent: Integrating large language models for interactive recommendations.
\newblock {\em arXiv preprint arXiv:2308.16505}, 2023.

\bibitem{wang2023enhancing}
Bing Wang, Xinnian Liang, Jian Yang, Hui Huang, Shuangzhi Wu, Peihao Wu, Lu~Lu, Zejun Ma, and Zhoujun Li.
\newblock Enhancing large language model with self-controlled memory framework.
\newblock {\em arXiv preprint arXiv:2304.13343}, 2023.

\bibitem{qian2307chatdev}
Chen Qian, Wei Liu, Hongzhang Liu, Nuo Chen, Yufan Dang, Jiahao Li, Cheng Yang, Weize Chen, Yusheng Su, Xin Cong, et~al.
\newblock Chatdev: Communicative agents for software development, 2024.
\newblock {\em URL https://arxiv. org/abs/2307}, 7924, 2024.

\bibitem{li2023metaagents}
Yuan Li, Yixuan Zhang, and Lichao Sun.
\newblock Metaagents: Simulating interactions of human behaviors for llm-based task-oriented coordination via collaborative generative agents.
\newblock {\em arXiv preprint arXiv:2310.06500}, 2023.

\bibitem{gao2023s}
Chen Gao, Xiaochong Lan, Zhihong Lu, Jinzhu Mao, Jinghua Piao, Huandong Wang, Depeng Jin, and Yong Li.
\newblock S$^{3}$: Social-network simulation system with large language model-empowered agents.
\newblock {\em arXiv preprint arXiv:2307.14984}, 2023.

\bibitem{li2023tradinggpt}
Yang Li, Yangyang Yu, Haohang Li, Zhi Chen, and Khaldoun Khashanah.
\newblock Tradinggpt: Multi-agent system with layered memory and distinct characters for enhanced financial trading performance.
\newblock {\em arXiv preprint arXiv:2309.03736}, 2023.

\bibitem{yen2024memolet}
Ryan Yen and Jian Zhao.
\newblock Memolet: Reifying the reuse of user-ai conversational memories.
\newblock In {\em Proceedings of the 37th Annual ACM Symposium on User Interface Software and Technology}, pages 1--22, 2024.

\bibitem{ko2024memreasoner}
Ching-Yun Ko, Sihui Dai, Payel Das, Georgios Kollias, Subhajit Chaudhury, and Aurelie Lozano.
\newblock Memreasoner: A memory-augmented llm architecture for multi-hop reasoning.
\newblock In {\em The First Workshop on System-2 Reasoning at Scale, NeurIPS'24}, 2024.

\bibitem{he2024madial}
Junqing He, Liang Zhu, Rui Wang, Xi~Wang, Reza Haffari, and Jiaxing Zhang.
\newblock Madial-bench: Towards real-world evaluation of memory-augmented dialogue generation.
\newblock {\em arXiv preprint arXiv:2409.15240}, 2024.

\bibitem{maharana2024evaluating}
Adyasha Maharana, Dong-Ho Lee, Sergey Tulyakov, Mohit Bansal, Francesco Barbieri, and Yuwei Fang.
\newblock Evaluating very long-term conversational memory of llm agents.
\newblock {\em arXiv preprint arXiv:2402.17753}, 2024.

\bibitem{zhang2024memsim}
Zeyu Zhang, Quanyu Dai, Luyu Chen, Zeren Jiang, Rui Li, Jieming Zhu, Xu~Chen, Yi~Xie, Zhenhua Dong, and Ji-Rong Wen.
\newblock Memsim: A bayesian simulator for evaluating memory of llm-based personal assistants.
\newblock {\em arXiv preprint arXiv:2409.20163}, 2024.

\bibitem{wu2025interpersonal}
Bowen Wu, Wenqing Wang, Haoran Li, Ying Li, Jingsong Yu, and Baoxun Wang.
\newblock Interpersonal memory matters: A new task for proactive dialogue utilizing conversational history.
\newblock {\em arXiv preprint arXiv:2503.05150}, 2025.

\bibitem{xu-etal-2022-beyond}
Jing Xu, Arthur Szlam, and Jason Weston.
\newblock Beyond goldfish memory: Long-term open-domain conversation.
\newblock In Smaranda Muresan, Preslav Nakov, and Aline Villavicencio, editors, {\em Proceedings of the 60th Annual Meeting of the Association for Computational Linguistics (Volume 1: Long Papers)}, pages 5180--5197, Dublin, Ireland, May 2022. Association for Computational Linguistics.

\bibitem{xue2025mmrc}
Haochen Xue, Feilong Tang, Ming Hu, Yexin Liu, Qidong Huang, Yulong Li, Chengzhi Liu, Zhongxing Xu, Chong Zhang, Chun-Mei Feng, et~al.
\newblock Mmrc: A large-scale benchmark for understanding multimodal large language model in real-world conversation.
\newblock {\em arXiv preprint arXiv:2502.11903}, 2025.

\bibitem{grauman2022ego4d}
Kristen Grauman, Andrew Westbury, Eugene Byrne, Zachary Chavis, Antonino Furnari, Rohit Girdhar, Jackson Hamburger, Hao Jiang, Miao Liu, Xingyu Liu, et~al.
\newblock Ego4d: Around the world in 3,000 hours of egocentric video.
\newblock In {\em Proceedings of the IEEE/CVF conference on computer vision and pattern recognition}, pages 18995--19012, 2022.

\bibitem{kuratov2024babilong}
Yuri Kuratov, Aydar Bulatov, Petr Anokhin, Ivan Rodkin, Dmitry Sorokin, Artyom Sorokin, and Mikhail Burtsev.
\newblock Babilong: Testing the limits of llms with long context reasoning-in-a-haystack, 2024.

\bibitem{kuratov2024search}
Yuri Kuratov, Aydar Bulatov, Petr Anokhin, Dmitry Sorokin, Artyom Sorokin, and Mikhail Burtsev.
\newblock In search of needles in a 10m haystack: Recurrent memory finds what llms miss, 2024.

\bibitem{gim2024prompt}
In~Gim, Guojun Chen, Seung-seob Lee, Nikhil Sarda, Anurag Khandelwal, and Lin Zhong.
\newblock Prompt cache: Modular attention reuse for low-latency inference.
\newblock {\em Proceedings of Machine Learning and Systems}, 6:325--338, 2024.

\bibitem{anthropic2024contextual}
Anthropic.
\newblock Introducing contextual retrieval.
\newblock {\em anthropic.com}, September 2024.

\bibitem{shao2023character}
Yunfan Shao, Linyang Li, Junqi Dai, and Xipeng Qiu.
\newblock Character-llm: A trainable agent for role-playing.
\newblock {\em arXiv preprint arXiv:2310.10158}, 2023.

\bibitem{qian2024memorag}
Hongjin Qian, Peitian Zhang, Zheng Liu, Kelong Mao, and Zhicheng Dou.
\newblock Memorag: Moving towards next-gen rag via memory-inspired knowledge discovery.
\newblock {\em arXiv preprint arXiv:2409.05591}, 2024.

\bibitem{liu2025echo}
WenTao Liu, Ruohua Zhang, Aimin Zhou, Feng Gao, and JiaLi Liu.
\newblock Echo: A large language model with temporal episodic memory.
\newblock {\em arXiv preprint arXiv:2502.16090}, 2025.

\bibitem{kadavy2021digital}
David Kadavy.
\newblock {\em Digital Zettelkasten: Principles, Methods, \& Examples}.
\newblock Kadavy, Inc., 2021.

\bibitem{murre2015replication}
Jaap~MJ Murre and Joeri Dros.
\newblock Replication and analysis of ebbinghaus’ forgetting curve.
\newblock {\em PloS one}, 10(7):e0120644, 2015.

\bibitem{karpukhin2020dense}
Vladimir Karpukhin, Barlas Oguz, Sewon Min, Patrick~SH Lewis, Ledell Wu, Sergey Edunov, Danqi Chen, and Wen-tau Yih.
\newblock Dense passage retrieval for open-domain question answering.
\newblock In {\em EMNLP (1)}, pages 6769--6781, 2020.

\bibitem{johnson2019billion}
Jeff Johnson, Matthijs Douze, and Herv{\'e} J{\'e}gou.
\newblock Billion-scale similarity search with gpus.
\newblock {\em IEEE Transactions on Big Data}, 7(3):535--547, 2019.

\bibitem{wang2024knowledge}
Song Wang, Yaochen Zhu, Haochen Liu, Zaiyi Zheng, Chen Chen, and Jundong Li.
\newblock Knowledge editing for large language models: A survey.
\newblock {\em ACM Computing Surveys}, 57(3):1--37, 2024.

\bibitem{han2024parameter}
Zeyu Han, Chao Gao, Jinyang Liu, Jeff Zhang, and Sai~Qian Zhang.
\newblock Parameter-efficient fine-tuning for large models: A comprehensive survey.
\newblock {\em arXiv preprint arXiv:2403.14608}, 2024.

\bibitem{hao2023reasoning}
Shibo Hao, Yi~Gu, Haodi Ma, Joshua~Jiahua Hong, Zhen Wang, Daisy~Zhe Wang, and Zhiting Hu.
\newblock Reasoning with language model is planning with world model.
\newblock {\em arXiv preprint arXiv:2305.14992}, 2023.

\bibitem{shinn2024reflexion}
Noah Shinn, Federico Cassano, Ashwin Gopinath, Karthik Narasimhan, and Shunyu Yao.
\newblock Reflexion: Language agents with verbal reinforcement learning.
\newblock {\em Advances in Neural Information Processing Systems}, 36, 2024.

\bibitem{christakopoulou2024agents}
Konstantina Christakopoulou, Shibl Mourad, and Maja Matari{\'c}.
\newblock Agents thinking fast and slow: A talker-reasoner architecture.
\newblock {\em arXiv preprint arXiv:2410.08328}, 2024.

\bibitem{ruan2023tptu}
Jingqing Ruan, Yihong Chen, Bin Zhang, Zhiwei Xu, Tianpeng Bao, Guoqing Du, Shiwei Shi, Hangyu Mao, Ziyue Li, Xingyu Zeng, et~al.
\newblock Tptu: large language model-based ai agents for task planning and tool usage.
\newblock {\em arXiv preprint arXiv:2308.03427}, 2023.

\bibitem{yang2024buffer}
Ling Yang, Zhaochen Yu, Tianjun Zhang, Shiyi Cao, Minkai Xu, Wentao Zhang, Joseph~E Gonzalez, and Bin Cui.
\newblock Buffer of thoughts: Thought-augmented reasoning with large language models.
\newblock {\em arXiv preprint arXiv:2406.04271}, 2024.

\bibitem{wang2024agent}
Zora~Zhiruo Wang, Jiayuan Mao, Daniel Fried, and Graham Neubig.
\newblock Agent workflow memory.
\newblock {\em arXiv preprint arXiv:2409.07429}, 2024.

\bibitem{liu2023think}
Lei Liu, Xiaoyan Yang, Yue Shen, Binbin Hu, Zhiqiang Zhang, Jinjie Gu, and Guannan Zhang.
\newblock Think-in-memory: Recalling and post-thinking enable llms with long-term memory.
\newblock {\em arXiv preprint arXiv:2311.08719}, 2023.

\bibitem{zhu2023ghost}
Xizhou Zhu, Yuntao Chen, Hao Tian, Chenxin Tao, Weijie Su, Chenyu Yang, Gao Huang, Bin Li, Lewei Lu, Xiaogang Wang, et~al.
\newblock Ghost in the minecraft: Generally capable agents for open-world environments via large language models with text-based knowledge and memory.
\newblock {\em arXiv preprint arXiv:2305.17144}, 2023.

\bibitem{wang2023voyager}
Guanzhi Wang, Yuqi Xie, Yunfan Jiang, Ajay Mandlekar, Chaowei Xiao, Yuke Zhu, Linxi Fan, and Anima Anandkumar.
\newblock Voyager: An open-ended embodied agent with large language models.
\newblock {\em arXiv preprint arXiv:2305.16291}, 2023.

\bibitem{yao2023retroformer}
Weiran Yao, Shelby Heinecke, Juan~Carlos Niebles, Zhiwei Liu, Yihao Feng, Le~Xue, Rithesh Murthy, Zeyuan Chen, Jianguo Zhang, Devansh Arpit, et~al.
\newblock Retroformer: Retrospective large language agents with policy gradient optimization.
\newblock {\em arXiv preprint arXiv:2308.02151}, 2023.

\bibitem{zhao2024expel}
Andrew Zhao, Daniel Huang, Quentin Xu, Matthieu Lin, Yong-Jin Liu, and Gao Huang.
\newblock Expel: Llm agents are experiential learners.
\newblock In {\em Proceedings of the AAAI Conference on Artificial Intelligence}, volume~38, pages 19632--19642, 2024.

\bibitem{zheng2023synapse}
Longtao Zheng, Rundong Wang, Xinrun Wang, and Bo~An.
\newblock Synapse: Trajectory-as-exemplar prompting with memory for computer control.
\newblock In {\em The Twelfth International Conference on Learning Representations}, 2023.

\bibitem{hong2023metagpt}
Sirui Hong, Xiawu Zheng, Jonathan Chen, Yuheng Cheng, Jinlin Wang, Ceyao Zhang, Zili Wang, Steven Ka~Shing Yau, Zijuan Lin, Liyang Zhou, et~al.
\newblock Metagpt: Meta programming for multi-agent collaborative framework.
\newblock {\em arXiv preprint arXiv:2308.00352}, 2023.

\bibitem{michelman2025enhancing}
Julie Michelman, Nasrin Baratalipour, and Matthew Abueg.
\newblock Enhancing reasoning with collaboration and memory.
\newblock {\em arXiv preprint arXiv:2503.05944}, 2025.

\bibitem{wang2025m+}
Yu~Wang, Dmitry Krotov, Yuanzhe Hu, Yifan Gao, Wangchunshu Zhou, Julian McAuley, Dan Gutfreund, Rogerio Feris, and Zexue He.
\newblock M+: Extending memoryllm with scalable long-term memory.
\newblock {\em arXiv preprint arXiv:2502.00592}, 2025.

\bibitem{zeng2023lookupffn}
Zhanpeng Zeng, Michael Davies, Pranav Pulijala, Karthikeyan Sankaralingam, and Vikas Singh.
\newblock Lookupffn: making transformers compute-lite for cpu inference.
\newblock In {\em International Conference on Machine Learning}, pages 40707--40718. PMLR, 2023.

\bibitem{liu2025chunkkv}
Xiang Liu, Zhenheng Tang, Peijie Dong, Zeyu Li, Bo~Li, Xuming Hu, and Xiaowen Chu.
\newblock Chunkkv: Semantic-preserving kv cache compression for efficient long-context llm inference.
\newblock {\em arXiv preprint arXiv:2502.00299}, 2025.

\bibitem{kwon2023efficient}
Woosuk Kwon, Zhuohan Li, Siyuan Zhuang, Ying Sheng, Lianmin Zheng, Cody~Hao Yu, Joseph Gonzalez, Hao Zhang, and Ion Stoica.
\newblock Efficient memory management for large language model serving with pagedattention.
\newblock In {\em Proceedings of the 29th Symposium on Operating Systems Principles}, pages 611--626, 2023.

\bibitem{wu2023fast}
Bingyang Wu, Yinmin Zhong, Zili Zhang, Shengyu Liu, Fangyue Liu, Yuanhang Sun, Gang Huang, Xuanzhe Liu, and Xin Jin.
\newblock Fast distributed inference serving for large language models.
\newblock {\em arXiv preprint arXiv:2305.05920}, 2023.

\bibitem{xiao2023efficient}
Guangxuan Xiao, Yuandong Tian, Beidi Chen, Song Han, and Mike Lewis.
\newblock Efficient streaming language models with attention sinks.
\newblock {\em arXiv preprint arXiv:2309.17453}, 2023.

\bibitem{yu2022orca}
Gyeong-In Yu, Joo~Seong Jeong, Geon-Woo Kim, Soojeong Kim, and Byung-Gon Chun.
\newblock Orca: A distributed serving system for $\{$Transformer-Based$\}$ generative models.
\newblock In {\em 16th USENIX Symposium on Operating Systems Design and Implementation (OSDI 22)}, pages 521--538, 2022.

\bibitem{zhong2024distserve}
Yinmin Zhong, Shengyu Liu, Junda Chen, Jianbo Hu, Yibo Zhu, Xuanzhe Liu, Xin Jin, and Hao Zhang.
\newblock $\{$DistServe$\}$: Disaggregating prefill and decoding for goodput-optimized large language model serving.
\newblock In {\em 18th USENIX Symposium on Operating Systems Design and Implementation (OSDI 24)}, pages 193--210, 2024.

\bibitem{dettmers2022gpt3}
Tim Dettmers, Mike Lewis, Younes Belkada, and Luke Zettlemoyer.
\newblock Gpt3. int8 (): 8-bit matrix multiplication for transformers at scale.
\newblock {\em Advances in neural information processing systems}, 35:30318--30332, 2022.

\bibitem{ge2023model}
Suyu Ge, Yunan Zhang, Liyuan Liu, Minjia Zhang, Jiawei Han, and Jianfeng Gao.
\newblock Model tells you what to discard: Adaptive kv cache compression for llms.
\newblock {\em arXiv preprint arXiv:2310.01801}, 2023.

\bibitem{li2020train}
Zhuohan Li, Eric Wallace, Sheng Shen, Kevin Lin, Kurt Keutzer, Dan Klein, and Joey Gonzalez.
\newblock Train big, then compress: Rethinking model size for efficient training and inference of transformers.
\newblock In {\em International Conference on machine learning}, pages 5958--5968. PMLR, 2020.

\bibitem{liu2023scissorhands}
Zichang Liu, Aditya Desai, Fangshuo Liao, Weitao Wang, Victor Xie, Zhaozhuo Xu, Anastasios Kyrillidis, and Anshumali Shrivastava.
\newblock Scissorhands: Exploiting the persistence of importance hypothesis for llm kv cache compression at test time.
\newblock {\em Advances in Neural Information Processing Systems}, 36:52342--52364, 2023.

\bibitem{zhang2023h2o}
Zhenyu Zhang, Ying Sheng, Tianyi Zhou, Tianlong Chen, Lianmin Zheng, Ruisi Cai, Zhao Song, Yuandong Tian, Christopher R{\'e}, Clark Barrett, et~al.
\newblock H2o: Heavy-hitter oracle for efficient generative inference of large language models.
\newblock {\em Advances in Neural Information Processing Systems}, 36:34661--34710, 2023.

\bibitem{qin2024mooncake}
Ruoyu Qin, Zheming Li, Weiran He, Mingxing Zhang, Yongwei Wu, Weimin Zheng, and Xinran Xu.
\newblock Mooncake: A kvcache-centric disaggregated architecture for llm serving.
\newblock {\em arXiv preprint arXiv:2407.00079}, 2024.

\bibitem{hu2024memserve}
Cunchen Hu, Heyang Huang, Junhao Hu, Jiang Xu, Xusheng Chen, Tao Xie, Chenxi Wang, Sa~Wang, Yungang Bao, Ninghui Sun, et~al.
\newblock Memserve: Context caching for disaggregated llm serving with elastic memory pool.
\newblock {\em arXiv preprint arXiv:2406.17565}, 2024.

\bibitem{recasens2024towards}
Pol~G Recasens, Yue Zhu, Chen Wang, Eun~Kyung Lee, Olivier Tardieu, Alaa Youssef, Jordi Torres, and Josep~Ll Berral.
\newblock Towards pareto optimal throughput in small language model serving.
\newblock In {\em Proceedings of the 4th Workshop on Machine Learning and Systems}, pages 144--152, 2024.

\bibitem{chen2025impress}
Weijian Chen, Shuibing He, Haoyang Qu, Ruidong Zhang, Siling Yang, Ping Chen, Yi~Zheng, Baoxing Huai, and Gang Chen.
\newblock $\{$IMPRESS$\}$: An $\{$Importance-Informed$\}$$\{$Multi-Tier$\}$ prefix $\{$KV$\}$ storage system for large language model inference.
\newblock In {\em 23rd USENIX Conference on File and Storage Technologies (FAST 25)}, pages 187--201, 2025.

\bibitem{li2025adaserve}
Zikun Li, Zhuofu Chen, Remi Delacourt, Gabriele Oliaro, Zeyu Wang, Qinghan Chen, Shuhuai Lin, April Yang, Zhihao Zhang, Zhuoming Chen, et~al.
\newblock Adaserve: Slo-customized llm serving with fine-grained speculative decoding.
\newblock {\em arXiv preprint arXiv:2501.12162}, 2025.

\bibitem{zhao2025mpic}
Shiju Zhao, Junhao Hu, Rongxiao Huang, Jiaqi Zheng, and Guihai Chen.
\newblock Mpic: Position-independent multimodal context caching system for efficient mllm serving.
\newblock {\em arXiv preprint arXiv:2502.01960}, 2025.

\bibitem{li2024intelllm}
TingLong Li and Qiuyu Shao.
\newblock Intel{LLM}: Little hints make a big difference for {LLM} {KV} cache compression, 2024.

\bibitem{pope2023efficiently}
Reiner Pope, Sholto Douglas, Aakanksha Chowdhery, Jacob Devlin, James Bradbury, Jonathan Heek, Kefan Xiao, Shivani Agrawal, and Jeff Dean.
\newblock Efficiently scaling transformer inference.
\newblock {\em Proceedings of Machine Learning and Systems}, 5:606--624, 2023.

\bibitem{liu2023cachegen}
Yuhan Liu, Hanchen Li, Kuntai Du, Jiayi Yao, Yihua Cheng, Yuyang Huang, Shan Lu, Michael Maire, Henry Hoffmann, Ari Holtzman, et~al.
\newblock Cachegen: Fast context loading for language model applications.
\newblock {\em CoRR}, 2023.

\bibitem{ye2024chunkattention}
Lu~Ye, Ze~Tao, Yong Huang, and Yang Li.
\newblock Chunkattention: Efficient self-attention with prefix-aware kv cache and two-phase partition.
\newblock {\em arXiv preprint arXiv:2402.15220}, 2024.

\bibitem{jin2024ragcache}
Chao Jin, Zili Zhang, Xuanlin Jiang, Fangyue Liu, Xin Liu, Xuanzhe Liu, and Xin Jin.
\newblock Ragcache: Efficient knowledge caching for retrieval-augmented generation.
\newblock {\em arXiv preprint arXiv:2404.12457}, 2024.

\bibitem{zheng2024efficiently}
Lianmin Zheng, Liangsheng Yin, Zhiqiang Xie, Chuyue Sun, Jeff Huang, Cody~Hao Yu, Shiyi Cao, Christos Kozyrakis, Ion Stoica, Joseph~E. Gonzalez, Clark Barrett, and Ying Sheng.
\newblock Sglang: Efficient execution of structured language model programs, 2024.

\bibitem{feng2024ada}
Yuan Feng, Junlin Lv, Yukun Cao, Xike Xie, and S~Kevin Zhou.
\newblock Ada-kv: Optimizing kv cache eviction by adaptive budget allocation for efficient llm inference.
\newblock {\em arXiv preprint arXiv:2407.11550}, 2024.

\bibitem{gao2024fast}
Shiwei Gao, Youmin Chen, and Jiwu Shu.
\newblock Fast state restoration in llm serving with hcache.
\newblock {\em arXiv preprint arXiv:2410.05004}, 2024.

\bibitem{jin2024compute}
Shuowei Jin, Xueshen Liu, Qingzhao Zhang, and Z~Morley Mao.
\newblock Compute or load kv cache? why not both?
\newblock {\em arXiv preprint arXiv:2410.03065}, 2024.

\bibitem{hu2024epic}
Junhao Hu, Wenrui Huang, Haoyi Wang, Weidong Wang, Tiancheng Hu, Qin Zhang, Hao Feng, Xusheng Chen, Yizhou Shan, and Tao Xie.
\newblock Epic: Efficient position-independent context caching for serving large language models.
\newblock {\em arXiv preprint arXiv:2410.15332}, 2024.

\bibitem{zhu2024relayattention}
Lei Zhu, Xinjiang Wang, Wayne Zhang, and Rynson~WH Lau.
\newblock Relayattention for efficient large language model serving with long system prompts.
\newblock {\em arXiv preprint arXiv:2402.14808}, 2024.

\bibitem{pan2024marconi}
Rui Pan, Zhuang Wang, Zhen Jia, Can Karakus, Luca Zancato, Tri Dao, Yida Wang, and Ravi Netravali.
\newblock Marconi: Prefix caching for the era of hybrid llms.
\newblock {\em arXiv preprint arXiv:2411.19379}, 2024.

\bibitem{quinn2024accelerating}
Derrick Quinn, Mohammad Nouri, Neel Patel, John Salihu, Alireza Salemi, Sukhan Lee, Hamed Zamani, and Mohammad Alian.
\newblock Accelerating retrieval-augmented generation.
\newblock {\em arXiv preprint arXiv:2412.15246}, 2024.

\bibitem{zhu2025fastcache}
Jianian Zhu, Hang Wu, Haojie Wang, Yinghui Li, Biao Hou, Ruixuan Li, and Jidong Zhai.
\newblock Fastcache: Optimizing multimodal llm serving through lightweight kv-cache compression framework.
\newblock {\em arXiv preprint arXiv:2503.08461}, 2025.

\bibitem{agarwal2025cache}
Shubham Agarwal, Sai Sundaresan, Subrata Mitra, Debabrata Mahapatra, Archit Gupta, Rounak Sharma, Nirmal~Joshua Kapu, Tong Yu, and Shiv Saini.
\newblock Cache-craft: Managing chunk-caches for efficient retrieval-augmented generation.
\newblock {\em arXiv preprint arXiv:2502.15734}, 2025.

\bibitem{yang2025kvlink}
Jingbo Yang, Bairu Hou, Wei Wei, Yujia Bao, and Shiyu Chang.
\newblock Kvlink: Accelerating large language models via efficient kv cache reuse.
\newblock {\em arXiv preprint arXiv:2502.16002}, 2025.

\bibitem{ray2024ragserve}
Siddhant Ray, Rui Pan, Zhuohan Gu, Kuntai Du, Ganesh Ananthanarayanan, Ravi Netravali, and Junchen Jiang.
\newblock Ragserve: Fast quality-aware rag systems with configuration adaptation.
\newblock {\em arXiv preprint arXiv:2412.10543}, 2024.

\bibitem{kumaribumblebee}
Lilly Kumari, Shengjie Wang, Tianyi Zhou, Nikhil Sarda, Anthony Rowe, and Jeff Bilmes.
\newblock Bumblebee: Dynamic kv-cache streaming submodular summarization for infinite-context transformers.
\newblock In {\em First Conference on Language Modeling}, 2024.

\bibitem{wu2022memorizing}
Yuhuai Wu, Markus~N Rabe, DeLesley Hutchins, and Christian Szegedy.
\newblock Memorizing transformers.
\newblock {\em arXiv preprint arXiv:2203.08913}, 2022.

\bibitem{tworkowski2024focused}
Szymon Tworkowski, Konrad Staniszewski, Miko{\l}aj Pacek, Yuhuai Wu, Henryk Michalewski, and Piotr Mi{\l}o{\'s}.
\newblock Focused transformer: Contrastive training for context scaling.
\newblock {\em Advances in Neural Information Processing Systems}, 36, 2024.

\bibitem{tack2024online}
Jihoon Tack, Jaehyung Kim, Eric Mitchell, Jinwoo Shin, Yee~Whye Teh, and Jonathan~Richard Schwarz.
\newblock Online adaptation of language models with a memory of amortized contexts.
\newblock {\em arXiv preprint arXiv:2403.04317}, 2024.

\bibitem{wang2024memoryllm}
Yu~Wang, Yifan Gao, Xiusi Chen, Haoming Jiang, Shiyang Li, Jingfeng Yang, Qingyu Yin, Zheng Li, Xian Li, Bing Yin, et~al.
\newblock Memoryllm: Towards self-updatable large language models.
\newblock {\em arXiv preprint arXiv:2402.04624}, 2024.

\bibitem{wang2024wise}
Peng Wang, Zexi Li, Ningyu Zhang, Ziwen Xu, Yunzhi Yao, Yong Jiang, Pengjun Xie, Fei Huang, and Huajun Chen.
\newblock Wise: Rethinking the knowledge memory for lifelong model editing of large language models.
\newblock {\em arXiv preprint arXiv:2405.14768}, 2024.

\bibitem{wang2023augmenting}
Weizhi Wang, Li~Dong, Hao Cheng, Xiaodong Liu, Xifeng Yan, Jianfeng Gao, and Furu Wei.
\newblock Augmenting language models with long-term memory.
\newblock {\em Advances in Neural Information Processing Systems}, 36:74530--74543, 2023.

\bibitem{kanglm2}
Jikun Kang, Wenqi Wu, Filippos Christianos, Alex~James Chan, Fraser~David Greenlee, George Thomas, Marvin Purtorab, and Andrew Toulis.
\newblock Lm2: Large memory models for long context reasoning.
\newblock In {\em Workshop on Reasoning and Planning for Large Language Models}, 2025.

\bibitem{behrouz2024titans}
Ali Behrouz, Peilin Zhong, and Vahab Mirrokni.
\newblock Titans: Learning to memorize at test time.
\newblock {\em arXiv preprint arXiv:2501.00663}, 2024.

\bibitem{vaswani2017attention}
Ashish Vaswani, Noam Shazeer, Niki Parmar, Jakob Uszkoreit, Llion Jones, Aidan~N Gomez, {\L}ukasz Kaiser, and Illia Polosukhin.
\newblock Attention is all you need.
\newblock {\em Advances in neural information processing systems}, 30, 2017.

\bibitem{zulfikar2024memoro}
Wazeer~Deen Zulfikar, Samantha Chan, and Pattie Maes.
\newblock Memoro: Using large language models to realize a concise interface for real-time memory augmentation.
\newblock In {\em Proceedings of the 2024 CHI Conference on Human Factors in Computing Systems}, pages 1--18, 2024.

\bibitem{wang2025unveiling}
Bo~Wang, Weiyi He, Pengfei He, Shenglai Zeng, Zhen Xiang, Yue Xing, and Jiliang Tang.
\newblock Unveiling privacy risks in llm agent memory.
\newblock {\em arXiv preprint arXiv:2502.13172}, 2025.

\end{thebibliography}

\end{document}